\documentclass[preprint]{aastex}
\usepackage{amsmath}
\usepackage{graphicx}
\usepackage{epstopdf}
\shortauthors{Bachelet et al.}
\shorttitle{MOA~2010-BLG-477Lb}

\newcommand{\bpi}{\mbox{\boldmath $\pi$}}
\newcommand\ltsima{$\; \buildrel <\over\sim \;$}
\newcommand\simlt{\lower.5ex\hbox{\ltsima}}
\newcommand\gtsima{$\; \buildrel >\over\sim \;$}
\newcommand\simgt{\lower.5ex\hbox{\gtsima}}

\begin{document}
\title{MOA~2010-BLG-477Lb: constraining the mass of a microlensing planet from microlensing parallax, orbital motion and detection of blended light}

\author{
E.~Bachelet\altaffilmark{2},
I.-G.~Shin\altaffilmark{3},
C.~Han\altaffilmark{1,3},
P.~Fouqu\'e\altaffilmark{2},
A.~Gould\altaffilmark{4},
J.W.~Menzies\altaffilmark{5},
J.-P.~Beaulieu\altaffilmark{36},
D.P.~Bennett\altaffilmark{14},
I.A.~Bond\altaffilmark{6},
Subo~Dong\altaffilmark{38,4},
D.~Heyrovsk\'y\altaffilmark{7},
J.-B.~Marquette\altaffilmark{36},
J.~Marshall\altaffilmark{4},
J.~Skowron\altaffilmark{4},
R.A.~Street\altaffilmark{8},
T.~Sumi\altaffilmark{9,34},
A.~Udalski\altaffilmark{10},
\and
L.~Abe\altaffilmark{28},
K.~Agabi\altaffilmark{28},
M.D.~Albrow\altaffilmark{12},
W.~Allen\altaffilmark{30},
E.~Bertin\altaffilmark{36},
M.~Bos\altaffilmark{31},
D.M.~Bramich\altaffilmark{17},
J.~Chavez\altaffilmark{96},
G.W.~Christie\altaffilmark{26},
A.A.~Cole\altaffilmark{20},
N.~Crouzet\altaffilmark{28},
S.~Dieters\altaffilmark{20},
M.~Dominik\altaffilmark{25,666},
J.~Drummond\altaffilmark{32},
J.~Greenhill\altaffilmark{20},
T.~Guillot\altaffilmark{28},
C.B.~Henderson\altaffilmark{4},
F.V.~Hessman\altaffilmark{82},
K.~Horne\altaffilmark{25},
M.~Hundertmark\altaffilmark{25,82},
J.A.~Johnson\altaffilmark{4},
U.G.~J{\o}rgensen\altaffilmark{35,42},
R.~Kandori\altaffilmark{39},
C.~Liebig\altaffilmark{25,37},
D.~M\'ekarnia\altaffilmark{28},
J.~McCormick\altaffilmark{43},
D.~Moorhouse\altaffilmark{44},
T.~Nagayama\altaffilmark{40},
D.~Nataf\altaffilmark{4},
T.~Natusch\altaffilmark{26,27},
S.~Nishiyama\altaffilmark{41},
J.-P.~Rivet\altaffilmark{28},
K.C.~Sahu\altaffilmark{29},
Y.~Shvartzvald\altaffilmark{45},
G.~Thornley\altaffilmark{44},
A.R.~Tomczak\altaffilmark{46},
Y.~Tsapras\altaffilmark{8,33},
J.C.~Yee\altaffilmark{4},
\and 
V.~Batista\altaffilmark{4,36},
C.S.~Bennett\altaffilmark{95},
S.~Brillant\altaffilmark{18},
J.A.R.~Caldwell\altaffilmark{96},
A.~Cassan\altaffilmark{36},
E.~Corrales\altaffilmark{36},
C.~Coutures\altaffilmark{36},
D.~Dominis Prester\altaffilmark{97},
J.~Donatowicz\altaffilmark{98},
D.~Kubas\altaffilmark{18,36},
R.~Martin\altaffilmark{99},
A.~Williams\altaffilmark{99},
M.~Zub\altaffilmark{37}
(The PLANET Collaboration),\\
L.~Andrade de Almeida\altaffilmark{49},
D.L.~DePoy\altaffilmark{46},
B.S.~Gaudi\altaffilmark{4},
L.-W.~Hung\altaffilmark{4},
F.~Jablonski\altaffilmark{49},
S.~Kaspi\altaffilmark{45},
N.~Klein\altaffilmark{45},
C.-U.~Lee\altaffilmark{76},
Y.~Lee\altaffilmark{50},
J.-R.~Koo\altaffilmark{76},
D.~Maoz\altaffilmark{45},
J.A.~Mu\~{n}oz\altaffilmark{48},
R.W.~Pogge\altaffilmark{4},
D.~Polishook\altaffilmark{45},
A.~Shporer\altaffilmark{8,23}
(The $\mu$FUN Collaboration),\\
F.~Abe\altaffilmark{34},
C.S.~Botzler\altaffilmark{11},
P.~Chote\altaffilmark{13},
M.~Freeman\altaffilmark{11},
A.~Fukui\altaffilmark{22},
K.~Furusawa\altaffilmark{34},
P.~Harris\altaffilmark{13},
Y.~Itow\altaffilmark{34},
S.~Kobara\altaffilmark{34},
C.H.~Ling\altaffilmark{6},
K.~Masuda\altaffilmark{34},
Y.~Matsubara\altaffilmark{34},
N.~Miyake\altaffilmark{34},
K.~Ohmori\altaffilmark{34},
K.~Ohnishi\altaffilmark{16},
N.J.~Rattenbury\altaffilmark{11},
To.~Saito\altaffilmark{21},
D.J.~Sullivan\altaffilmark{13},
D.~Suzuki\altaffilmark{9},
W.L.~Sweatman\altaffilmark{6},
P.J.~Tristram\altaffilmark{15},
K. Wada\altaffilmark{9},
P.C.M.~Yock\altaffilmark{11}
(The MOA Collaboration),\\
M.K.~Szyma{\'n}ski\altaffilmark{10},
I.~Soszy{\'n}ski\altaffilmark{10},
M.~Kubiak\altaffilmark{10},  
R.~Poleski\altaffilmark{10},
K.~Ulaczyk\altaffilmark{10},
G.~Pietrzy{\'n}ski\altaffilmark{10,65}, 
{\L}.~Wyrzykowski\altaffilmark{10,66}
(The OGLE Collaboration),\\
N.~Kains\altaffilmark{17,25},
C.~Snodgrass\altaffilmark{90},
I.A.~Steele\altaffilmark{91}
(The RoboNet Collaboration),\\
K.A.~Alsubai\altaffilmark{77},
V.~Bozza\altaffilmark{24},
P.~Browne\altaffilmark{25},
M.J.~Burgdorf\altaffilmark{83,84},
S.~Calchi Novati\altaffilmark{24,79},
P.~Dodds\altaffilmark{25},
S.~Dreizler\altaffilmark{82},
F.~Finet\altaffilmark{85},
T.~Gerner\altaffilmark{37},
S.~Hardis\altaffilmark{35},
K.~Harps{\o}e\altaffilmark{35,42},
T.C.~Hinse\altaffilmark{76},
E.~Kerins\altaffilmark{86},
L.~Mancini\altaffilmark{24,80},
M.~Mathiasen\altaffilmark{35},
M.T.~Penny\altaffilmark{86},
S.~Proft\altaffilmark{37},
S.~Rahvar\altaffilmark{87},
D.~Ricci\altaffilmark{85},
G.~Scarpetta\altaffilmark{24,81},
S.~Sch\"{a}fer\altaffilmark{82},
F.~Sch\"{o}nebeck\altaffilmark{37},
J.~Southworth\altaffilmark{88},
J.~Surdej\altaffilmark{85},
J.~Wambsganss\altaffilmark{37}\\
(The MiNDSTEp Consortium)
}

\altaffiltext{1}{corresponding author}
\altaffiltext{2}{IRAP, Universit\'e de Toulouse, CNRS, 14 Avenue Edouard Belin, 31400 Toulouse, France}
\altaffiltext{3}{Department of Physics, Chungbuk National University, Cheongju 361-763, Korea}
\altaffiltext{4}{Department of Astronomy, Ohio State University, 140 West 18th Avenue, Columbus, OH 43210, USA}
\altaffiltext{5}{South African Astronomical Observatory, P.O. Box 9 Observatory 7925, South Africa}
\altaffiltext{6}{Institute for Information and Mathematical Sciences, Massey University, Private Bag 102-904, Auckland 1330, New Zealand}
\altaffiltext{7}{Institute of Theoretical Physics, Charles University, V Hole\v{s}ovi\v{c}k\'{a}ch 2, 18000 Prague, Czech Republic}
\altaffiltext{8}{Las Cumbres Observatory Global Telescope Network, 6740B Cortona Dr, Goleta, CA 93117, USA}
\altaffiltext{9}{Department of Earth and Space Science, Graduate School of Science, Osaka University, 1-1 Machikaneyama-cho, Toyonaka, Osaka 560-0043, Japan }
\altaffiltext{10}{Warsaw University Observatory, Al. Ujazdowskie 4, 00-478 Warszawa, Poland}
\altaffiltext{11}{Department of Physics, University of Auckland, Private Bag 92-019, Auckland 1001, New Zealand}
\altaffiltext{12}{University of Canterbury, Department of Physics and Astronomy, Private Bag 4800, Christchurch 8020, New Zealand}
\altaffiltext{13}{School of Chemical and Physical Sciences, Victoria University, Wellington, New Zealand}
\altaffiltext{14}{Department of Physics, 225 Nieuwland Science Hall, University of Notre Dame, Notre Dame, IN 46556, USA}
\altaffiltext{15}{Mt. John University Observatory, P.O. Box 56, Lake Tekapo 8770, New Zealand}
\altaffiltext{16}{Nagano National College of Technology, Nagano 381-8550, Japan}
\altaffiltext{17}{European Southern Observatory, Karl-Schwarzschild-Str. 2, 85748 Garching bei M\"{u}nchen, Germany}
\altaffiltext{18}{European Southern Observatory, Casilla 19001, Vitacura 19, Santiago, Chile}
\altaffiltext{19}{Jodrell Bank Observatory, The University of Manchester, Macclesfield, Cheshire SK11 9DL, UK}
\altaffiltext{20}{University of Tasmania, School of Mathematics and Physics, Private Bag 37, Hobart, TAS 7001, Australia}
\altaffiltext{21}{Tokyo Metropolitan College of Aeronautics, Tokyo 116-8523, Japan}
\altaffiltext{22}{Okayama Astrophysical Observatory, National Astronomical Observatory, 3037-5 Honjo, Kamogata, Asakuchi, Okayama 719-0232, Japan}
\altaffiltext{23}{Department of Physics, University of California, Santa Barbara, CA 93106, USA}
\altaffiltext{24}{Universit\`{a} degli Studi di Salerno, Dipartimento di Fisica "E.R.~ Caianiello", Via S.~Allende, 84081 Baronissi (SA), Italy}
\altaffiltext{25}{SUPA, University of St Andrews, School of Physics \& Astronomy, North Haugh, St Andrews, KY16 9SS, UK}
\altaffiltext{26}{Auckland Observatory, Auckland, New Zealand}
\altaffiltext{27}{AUT University, Auckland, New Zealand}
\altaffiltext{28}{Laboratoire Fizeau, Observatoire de la C\^ote d'Azur, Boulevard de l'Observatoire, 06300 Nice, France}
\altaffiltext{29}{Space Telescope Science Institute, 3700 San Martin Drive, Baltimore, MD 21218, USA}
\altaffiltext{30}{Vintage Lane Observatory, Blenheim, New Zealand}
\altaffiltext{31}{Molehill Astronomical Observatory, North Shore, New Zealand}
\altaffiltext{32}{Possum Observatory, New Zealand}
\altaffiltext{33}{School of Physics and Astronomy, Queen Mary University of London, Mile End Road, London, E1 4NS}
\altaffiltext{34}{Solar-Terrestrial Environment Laboratory, Nagoya University, Nagoya, 464-8601, Japan}
\altaffiltext{35}{Niels Bohr Institutet, K{\o}benhavns Universitet, Juliane Maries Vej 30, 2100 K{\o}benhavn {\O}, Denmark}
\altaffiltext{36}{UPMC-CNRS, UMR 7095, Institut d'Astrophysique de Paris, 98bis boulevard Arago, F-75014 Paris, France}
\altaffiltext{37}{Astronomisches Rechen-Institut, Zentrum f\"{u}r Astronomie der Universit\"{a}t Heidelberg (ZAH), M\"{o}nchhofstr.\ 12-14, 69120 Heidelberg, Germany}
\altaffiltext{38}{Institute for Advanced Study, Einstein Drive, Princeton, NJ 08540, USA}
\altaffiltext{39}{National Astronomical Observatory, 2-21-1 Osawa, Mitaka, Tokyo 181-8588, Japan}
\altaffiltext{40}{Graduate School of Science, Nagoya University, Furo-cho, Chikusa-ku, Nagoya 464-8602, Japan}
\altaffiltext{41}{Extrasolar Planet Detection Project Office, National Astronomical Observatory of Japan (NAOJ), Osawa 2-21-1, Mitaka, Tokyo 181-8588, Japan}
\altaffiltext{42}{Centre for Star and Planet Formation, K{\o}benhavns Universitet, {\O}ster Voldgade 5-7, 1350 K{\o}benhavn {\O}, Denmark}
\altaffiltext{43}{Farm Cove Observatory, Centre for Backyard Astrophysics, Pakuranga, Auckland, New Zealand}
\altaffiltext{44}{Kumeu Observatory, Kumeu, New Zealand}
\altaffiltext{45}{School of Physics and Astronomy and Wise Observatory, Tel-Aviv University, Tel-Aviv 69978, Israel}
\altaffiltext{46}{Department of Physics and Astronomy, Texas A\&M University, College Station, Texas 77843-4242, USA}
\altaffiltext{48}{Departamento de Astronomi{\'a} y Astrof{\'i}sica, Universidad de Valencia, E-46100 Burjassot, Valencia, Spain}
\altaffiltext{49}{Divisao de Astrofisica, Instituto Nacional de Pesquisas Espaciais, Avenida dos Astronautas, 1758 Sao Jos\'e dos Campos, 12227-010 SP, Brasil}
\altaffiltext{50}{Department of Astronomy and Space Science, Chungnam University, Korea}
\altaffiltext{65}{Universidad de Concepci\'on, Departamento de Astronom\'{\i}a, Casilla 160--C, Concepci\'on, Chile}
\altaffiltext{66}{Institute of Astronomy, University of Cambridge, Madingley Road, Cambridge CB3 0HA, UK}
\altaffiltext{75}{Armagh Observatory, College Hill, Armagh, BT61 9DG, Northern Ireland, UK}
\altaffiltext{76}{Korea Astronomy and Space Science Institute, 776 Daedukdae-ro, Yuseong-gu, Daejeon 305-348, Republic of Korea}
\altaffiltext{77}{Qatar Foundation, P.O. Box 5825, Doha, Qatar}
\altaffiltext{79}{Istituto Internazionale per gli Alti Studi Scientifici (IIASS), Vietri Sul Mare (SA), Italy}
\altaffiltext{80}{Max Planck Institute for Astronomy, K\"{o}nigstuhl 17, 619117 Heidelberg, Germany}
\altaffiltext{81}{INFN, Gruppo Collegato di Salerno, Sezione di Napoli, Italy}
\altaffiltext{82}{Institut f\"{u}r Astrophysik, Georg-August-Universit\"{a}t, Friedrich-Hund-Platz 1, 37077 G\"{o}ttingen, Germany}
\altaffiltext{83}{Deutsches SOFIA Institut, Universit\"{a}t Stuttgart, Pfaffenwaldring 31, 70569 Stuttgart, Germany}
\altaffiltext{84}{SOFIA Science Center, NASA Ames Research Center, Mail Stop N211-3, Moffett Field CA 94035, USA}
\altaffiltext{85}{Institut d'Astrophysique et de G\'{e}ophysique, All\'{e}e du 6 Ao\^{u}t 17, Sart Tilman, B\^{a}t.\ B5c, 4000 Li\`{e}ge, Belgium}
\altaffiltext{86}{Jodrell Bank Centre for Astrophysics, University of Manchester, Oxford Road,Manchester, M13 9PL, UK}
\altaffiltext{87}{Department of Physics, Sharif University of Technology, P.~O.\ Box 11155--9161, Tehran, Iran}
\altaffiltext{88}{Astrophysics Group, Keele University, Staffordshire, ST5 5BG, UK}
\altaffiltext{90}{Max Planck Institute for Solar System Research, Max-Planck-Str. 2, 37191 Katlenburg-Lindau, Germany}
\altaffiltext{91}{Astrophysics Research Institute, Liverpool John Moores University, Liverpool CH41 1LD, UK}
\altaffiltext{95}{Department of Physics, Massachussets Institute of Technology, 77 Mass. Ave., Cambridge, MA 02139, USA}
\altaffiltext{96}{McDonald Observatory, 16120 St Hwy Spur 78 \#2, Fort Davis, Texas 79734, USA}
\altaffiltext{97}{Department of Physics, University of Rijeka, Omladinska 14, 51000 Rijeka, Croatia}
\altaffiltext{98}{Technische Universit\"{a}t Wien, Wieder Hauptst. 8-10, A-1040 Vienna, Austria}
\altaffiltext{99}{Perth Observatory, Walnut Road, Bickley, Perth 6076, WA, Australia}
\altaffiltext{666}{Royal Society University Research Fellow}


\begin{abstract}
Microlensing detections of cool planets are important for the construction of 
an unbiased sample to estimate the frequency of planets beyond the snow line, 
which is where giant planets are thought to form according to the core 
accretion theory of planet formation. In this paper, we report the discovery of 
a giant planet detected from the analysis of the light curve of a 
high-magnification microlensing event MOA~2010-BLG-477. The measured 
planet-star mass ratio is $q=(2.181\pm0.004)\times 10^{-3}$ and the projected 
separation is $s=1.1228\pm0.0006$ in units of the Einstein radius. The angular 
Einstein radius is unusually large $\theta_{\rm E}=1.38\pm 0.11$~mas. Combining 
this measurement with constraints on the ``microlens parallax'' and the lens 
flux, we can only limit the host mass to the range $0.13<M/M_\odot<1.0$. In this
particular case, the strong degeneracy between microlensing parallax and planet 
orbital motion prevents us from measuring more accurate host and planet masses.
However, we find that adding Bayesian priors from two effects (Galactic model 
and Keplerian orbit) each independently favors the upper end of this mass range,
yielding star and planet masses of $M_*=0.67^{+0.33}_{-0.13}\ M_\odot$ and 
$m_p=1.5^{+0.8}_{-0.3}\ M_{\rm JUP}$ at a distance of $D=2.3\pm0.6$~kpc, and 
with a semi-major axis of $a=2^{+3}_{-1}$~AU. Finally, we show that the lens 
mass can be determined from future high-resolution near-IR adaptive optics 
observations independently from two effects, photometric and astrometric.
\end{abstract}



\section{Introduction}

Gravitational microlensing is an important method to detect extrasolar planets
\citep{mao91,bennett08a,gaudi10}. The method is sensitive to planets not easily 
accessible to other methods, in particular cool and small planets at or beyond 
the snow line \citep{beaulieu06, bennett08b}, and free-floating planets
\citep{sumi11}. The snow line represents the location in the protoplanetary 
disk beyond which ices can exist \citep{lecar06,kennedy07,kennedy08} and thus 
the surface density of solids is highest \citep{lissauer87}. According to the 
core accretion theory of planet formation \citep{lissauer93}, the snow line 
plays a crucial role because giant planets are thought to form in the region 
immediately beyond the snow line. Therefore, microlensing planets can provide 
important constraints on planet formation theories, in particular by measuring
the mass function beyond the snow line \citep{gould10,sumi10,cassan12}.

A major component of current planetary microlensing experiments is being 
carried out in survey and follow-up mode, where survey experiments are 
conducted in order to maximize the event rate by monitoring a large area of the
sky one or several times per night, while follow-up experiments are focused on 
events alerted by survey observations to densely cover planet-induced 
perturbations. In this mode, high-magnification events are important targets 
for follow-up observations. This is because the source trajectories of these 
events always pass close to the central perturbation region and thus the 
sensitivity to planets is extremely high 
\citep{griest98,rhie00,rattenbury02,abe04,han09}. In addition, the time of the 
perturbation can be predicted in advance so that intensive follow-up 
observation can be prepared. This leads to an observational strategy of 
monitoring high-magnification events as intensively as possible, regardless of
whether or not they show evidence of planets. As a result, the strategy allows
one to construct an unbiased sample to derive the frequency of planets 
beyond snow line \citep{gould10}. For the alternative low-magnification channel 
of detection, see for instance \citet{sumi10}.

In this paper, we report the discovery of a giant planet detected from the 
analysis of the light curve of a high-magnification microlensing event 
MOA~2010-BLG-477. Due to the high magnification of the event, the perturbation 
was very densely covered, enabling us to place constraints on the physical 
parameters from the higher-order effects in the lensing light curve induced by 
finite source effects as well as the orbital motion of both the lens and the 
Earth. We provide the most probable physical parameters of the planetary system,
corresponding to a Jupiter-mass planet orbiting a K~dwarf at about 2~AU, 
the system lying at about 2~kpc from Earth.

\section{Observations}\label{sec:obs}

The event MOA~2010-BLG-477 at coordinates
(RA, DEC)=($18^{\rm h}06^{\rm m}07.44^{\rm s}, -31^{\circ}27\arcmin16.12\arcsec)$ 
(J2000.0), $(l, b) = (0.046^{\circ}, -5.095^{\circ})$ was detected and announced 
as a microlensing alert event by the Microlensing Observations in Astrophysics 
\citep{bond01,sumi03} collaboration on 2010 August 2 
($\rm HJD'=HJD-2450000=5410.9$) using the 1.8 m telescope of Mt.\ John 
Observatory in New Zealand. The event was also observed by the Optical 
Gravitational Lensing Experiment \citep{udalski03} using the 1.3m Warsaw 
telescope of Las Campanas Observatory in Chile. 

Real-time modeling based on the rising part of the light curve indicated 
that the event would reach very high magnification and enabled us to predict 
the time of the peak. Followed by this second alert, the peak of the light 
curve was densely covered by follow-up groups of the Probing Lensing Anomalies 
NETwork (PLANET) \citep{beaulieu06}, Microlensing Follow-Up Network ($\mu$FUN) 
\citep{gould06}, RoboNet \citep{tsapras09}, and Microlensing Network 
for the Detection of Small Terrestrial Exoplanets (MiNDSTEp) \citep{dominik10}. 
More than twenty telescopes were used for the follow-up observations, including 
PLANET 1.0m of South African Astronomical Observatory (SAAO) in South Africa, 
PLANET 1.0m of Mt.\ Canopus Observatory in Tasmania, Australia, PLANET 0.6m 
Perth Observatory in Australia, PLANET 0.4m ASTEP telescope at Dome~C, 
Antarctica, $\mu$FUN 1.3m SMARTS telescope of the Cerro Tololo Inter-American 
Observatory (CTIO), Chile, $\mu$FUN 0.4m of Auckland Observatory, $\mu$FUN 
0.36m of Farm Cove Observatory (FCO), $\mu$FUN 0.36m of Kumeu Observatory, 
$\mu$FUN 0.4m of Possum Observatory, $\mu$FUN 0.4m of Vintage Lane Observatory 
(VLO), $\mu$FUN 0.3m of Molehill Astronomical Observatory (MAO), all in New 
Zealand, $\mu$FUN 0.46m of Wise Observatory, Israel, $\mu$FUN 0.8m of Teide 
Observatory at Canary Islands (IAC), Spain, $\mu$FUN 0.6m of Pico dos Dias 
Observatory, Brazil, RoboNet 2.0m of Faulkes North (FTN) in Hawaii, U.S.A., 
RoboNet 2.0m of Faulkes South (FTS), Australia, RoboNet 2.0m of Liverpool (LT) 
at Canary Islands, Spain, MiNDSTEp 1.54 m Danish telescope at La Silla, Chile, 
and MiNDSTEp 1.2m Monet North telescope at McDonald Observatory, U.S.A..  

Please note that this paper reports the first microlensing observations from 
Antarctica. Unfortunately, the data quality of these pioneering observations 
was not high enough to be included into the models, but we give a short 
overview in Appendix~\ref{app:astep}. In order to better constrain the 
second-order effects, new observations were taken at the $\mu$FUN 1.3 m SMARTS 
telescope at CTIO during the 2011 campaign.

\begin{figure}[th]
\epsscale{1.0}
\plotone{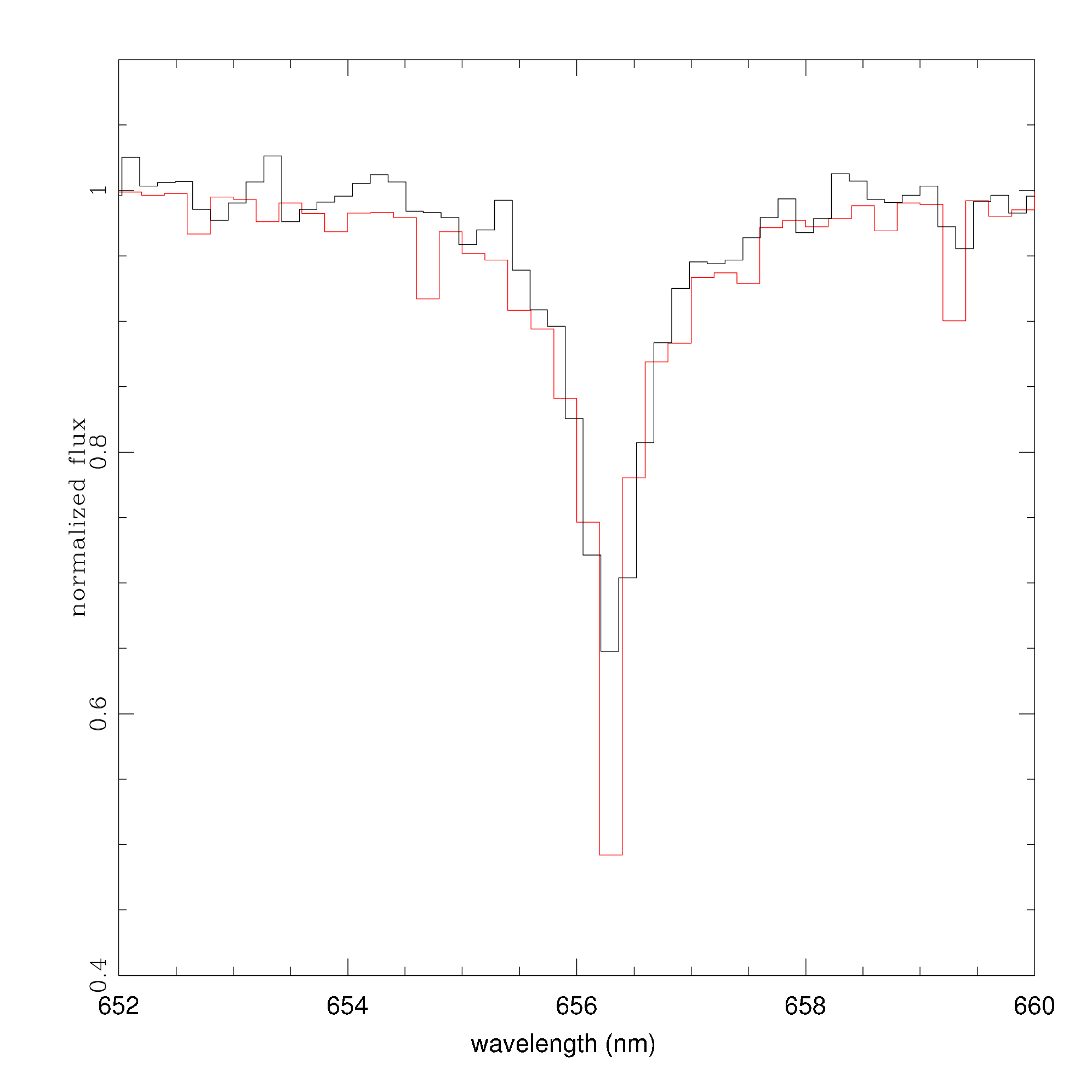}
\caption{\label{fig:spectrum}
Reduced spectrum with an overplotted synthetic spectrum
for the adopted source characteristics, namely $T_{\rm eff}=5950$~K, 
$\log g=4.0$ and solar metallicity.}
\end{figure}

To better characterize the lensed source star, spectroscopic observations 
were conducted near the peak of the event ($\rm HJD'=5422.5$) by using the 
B\&C spectrograph on the 2.5 m du Pont telescope at Las Campanas Observatory in 
Chile. The resolution was $R=1400$, corresponding to $\Delta\lambda=4.6$~\AA. A 
comparison between the observed spectrum and synthetic ones was conducted to 
derive the effective temperature, gravity and metallicity of the source star. 
The usable part of the spectrum is only $\sim1000$~\AA \  due to some scattered 
light issues with the instrument, particularly a problem for this faint star. 
We started by fitting standard stars using H$\alpha$, MgB, and NaD, but found 
that fits with H$\alpha$ alone resulted in the most accurate temperatures, so 
this is the diagnostic we used. The derived parameters of the source star are 
$T_{\rm eff}=5950 \pm 150$~K for $\log g=4.0$ and a solar metallicity. The 
corresponding fit to the H$\alpha$ line is shown in Figure~\ref{fig:spectrum}.

The data collected by the individual groups were initially reduced using 
various photometry codes developed by the individual groups. For the data sets 
from SAAO, FTS, Possum, Canopus, Perth, Danish and Monet North, we use 
photometry from rereductions obtained with the pySIS package, described in more
details in Appendix~\ref{app:pysis}. 

Figure~\ref{fig:lcurve} shows the light curve of the event. The event is highly 
magnified with a peak magnification $A\sim 400$. Outside of the region 
${\rm HJD'}=[5417,5425]$, the light curve is consistent with a standard 
single-lens curve \citep{paczynski86}. The perturbation is composed of two 
spikes at ${\rm HJD'}\sim 5420.4$ and 5420.9 and two bumps at 
${\rm HJD'}\sim 5421.0$ and 5422.4.

\begin{figure}[th]
\epsscale{1.0}
\plotone{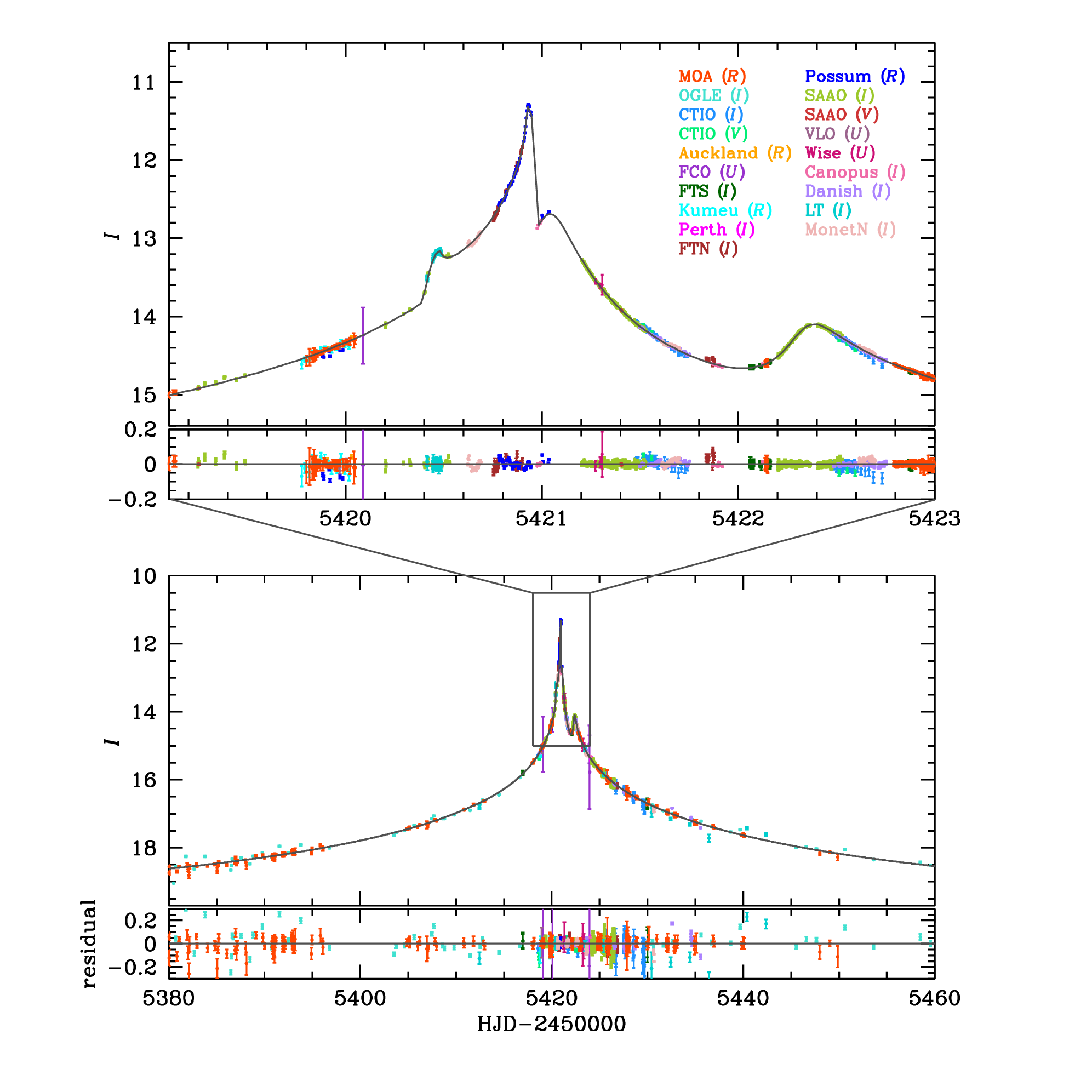}
\caption{\label{fig:lcurve}
Light curve of the microlensing event MOA~2010-BLG-477.  Data from different 
observatories are distinguished by different colors. The upper panel shows 
the enlargement of the region of perturbation near the peak of the light 
curve. Note that the model light curve corresponds to the best-fit solution 
of the model considering both parallax and orbital motion and the corresponding 
parameters are presented in Table~\ref{tab:one}.}
\end{figure}

\section{Strategic Overview}\label{sec:overview}

As in the great majority of planetary microlensing events, we are able to 
measure the "angular Einstein radius" (projected on the sky) $\theta_{\rm E}$, 
but not the projected Einstein radius (projected on the observer plane) 
$\tilde r_{\rm E}$. Consequently, it is challenging to estimate the physical 
parameters of the lens. The two radius quantities are related to those 
\citep[see]{gould00b} by
\begin{equation}
\theta_{\rm E}^2 = \kappa\,\pi_{\rm rel}\,M;
\quad
\pi_{\rm E}^2 = \frac{\pi_{\rm rel}}{\kappa\,M};
\quad
\tilde r_{\rm E} = \frac{{\rm AU}}{\pi_{\rm E}}
\label{eqn:theta_e+pi_e}
\end{equation}
where $\kappa \equiv 4GM/{\rm AU}c^2 \sim 8.1439\,{\rm mas}\,M_\odot^{-1}$, $M$
is the lens mass in $M_\odot$, and $\pi_{\rm rel} = {\rm AU}(1/D_L-1/D_S)$ is the 
lens-source relative parallax ($D_L$ and $D_S$ are the lens and source 
distances). Since $\theta_{\rm E}$ is well measured, the product 
$\pi_{\rm rel}\,M$ is also well determined, but in the present case, the ratio 
$\pi_{\rm rel}/M$ is poorly constrained, and so it is difficult to estimate $M$
alone. In this section, we provide an overview of the various techniques that 
we use to place constraints on the individual quantities $\pi_{\rm rel}$ and $M$.

We will show that here $\theta_{\rm E}$ is large enough to enable a substantial 
constraint from upper limits on the lens flux. That is, from 
Equation~(\ref{eqn:theta_e+pi_e}), large $\theta_{\rm E}$ implies large $M$ 
or $\pi_{\rm rel}$, both of which lead to brighter lenses. This will lead to the 
unambiguous conclusion that the lens is nearby, $D_L\la 3\,$kpc, with mass
$M\la 1\,M_\odot$.  

The light curve of this event enables stronger constraints than is usually the 
case because we are able to obtain a measurement of one of the microlens
parallax component. The microlens parallax is actually a vector, $\bpi_{\rm E}$, 
with the magnitude given by Equation~(\ref{eqn:theta_e+pi_e}) and the direction 
by the lens-source relative motion \citep{gould94a}. One component of the 
parallax only weakly constrains the scalar $\pi_{\rm E}$, but it constrains the 
direction of proper motion, which will be very important for future 
observations (see below). Moreover, one parallax component actually does 
provide a robust lower limit on the mass $M\ga 0.13\,M_\odot$ and distance 
$D_L\ga 0.5\,$kpc.

To proceed further, we must apply a Bayesian analysis, but here as elsewhere we
are more fortunate than is typical. As always, there are Bayesian priors from a
Galactic model, and in this case these strongly prefer the upper end of the 
mass/distance range permitted by the light curve.

But in addition, the reason that only one dimension of $\bpi_{\rm E}$ can be 
robustly measured is that the other dimension is degenerate with orbital motion
of the planet \citep{batista11}. Thus, it is necessary to fit simultaneously
for both parallax and orbital motion. Bayesian priors on the orbital motion 
then {\it independently} also prefer the upper end of the allowed mass/distance
parameter space.

Finally, we predict that high-resolution imaging could measure the mass and 
distance of the lens through two independent effects, photometry and 
astrometry. The first is obvious: different mass/distance combinations yield 
different color/magnitude measurement. The main point here is that the large 
value of $\theta_{\rm E}$ virtually guarantees that the lens will be detectable.
The second is less obvious: different mass/distance combinations also predict
different {\it directions} of source-lens relative proper motion, which can be 
measured as the lens and source are separated over the next several years.

\section{Modeling}

\subsection{Treatment of photometric errors}

The photometric error bars of the various data sets from individual 
observatories are generally not accurate enough to be taken at face value. They
need a rescaling to reproduce the dispersion of contiguous data points in a
given night. At the same time, outliers must be identified and removed. These 
are important preliminary steps to modeling, because the weight of a given data
set to constrain the model depends on how large its error bars are, compared to
other data sets. We describe our adopted noise model and rescaling factors for 
each observatory data set in Appendix~\ref{app:rescaling}.

\subsection{Static Binary Model}

We first test a static binary lens model. The corresponding parameter set 
includes the three single lens parameters: the Einstein time scale, $t_{\rm E}$, 
the time of the closest lens-source approach, $t_0$, and the lens-source 
separation at that moment, $u_0$, and it also includes the three binary 
parameters: the mass ratio of the companion to its host star, $q$, the 
projected separation between the lens components in units of the Einstein 
radius, $s$, and the angle between the source trajectory and the binary axis, 
$\alpha$. Since the light curve exhibits caustic-crossing features, we need to 
consider the modification of magnifications caused by the finite-source effect 
\citep{nemiroff94,witt94,gould94b,bennett96,vermaak00}. This requires us to 
include an additional parameter: the normalized source radius
\begin{equation}
\rho_\star \equiv \frac{\theta_\star}{\theta_{\rm E}}
\label{eqn:rhodef}
\end{equation}
where $\theta_\star$ is the angular source radius. Evaluation of $\rho_\star$ 
from the model together with the measurement of $\theta_\star$ (see 
Section~\ref{sec:source}) will then yield $\theta_{\rm E}$ and thus the product 
$\pi_{\rm rel}\,M$ (see Equation~[\ref{eqn:theta_e+pi_e}]).

\subsection{Finite-source Effect}

Since the event MOA~2010-BLG-477 involves caustic crossings and approaches, 
one must compute magnification affected by the finite-source effect. 
Computation of finite magnifications is based on the numerical ray-shooting 
method. In this method, a large number of rays are uniformly shot from the 
image plane, bent according to the lens equation, and land on the source plane. 
The lens equation is represented by \citep{witt95}
\begin{equation} \label{eqn:eq3}
\zeta = z - \sum_{k=1}^2 \frac{m_k/M}{\bar{z}-\bar{z}_{L,k}},
\end{equation}
where $\zeta=\xi + i\eta$, $z_{L,k}=x_{L,k}+iy_{L,k}$, and $z=x+iy$ are the 
complex notations of the source, lens, and image positions, respectively, 
$\bar{z}$ denotes the complex conjugate of $z$, and $m_k$ are the masses of the
individual lens components, and $M$ is the total mass of the lens system. Then,
the magnification is computed as the ratio of the number density of rays on the
source plane to the density of the image plane.  For the initial search for 
solutions in the space of the grid parameters, we accelerate the computation by
using the ``map making'' method \citep{dong06}. In this method, a magnification
map is made for a given set of $(s,q)$ and then it is used to produce numerous 
light curves resulting from different source trajectories instead of 
re-shooting rays all over again.  We further accelerate the computation by 
using a semi-analytic hexadecapole approximation for finite-magnification 
computation \citep{pejcha09,gould08} in the region where the source location is
not very close to the caustic. In computing finite magnifications, we consider
the effect of limb darkening of the source star surface by modeling the 
specific intensity as \citep{milne21,an02}
\begin{equation} \label{eqn:eq4}
I_{\lambda} = \frac{F_{\lambda}}{\pi \theta^2 _{\star}} 
{ { \left[ 1- \Gamma_{\lambda} \left( 1- \frac{3}{2} \cos\phi \right) \right]} }
\end{equation}
where $\Gamma_{\lambda}$ is a limb-darkening coefficient (hereafter LDC), 
$F_{\lambda}$ is the total flux from the source star, $\phi$ is the angle 
between the direction toward the observer and the normal to the stellar 
surface. From the $\chi^2$ improvement we find that the limb-darkening effect 
is clearly detected. We compute the LDCs for Equation~\ref{eqn:eq4} as 
accurately as possible, including a proper treatment of the effect of 
extinction. We refer the interested reader to Appendix~\ref{app:lldc} for 
details.

\subsection{Microlensing Parallax and Planet Orbital Motion}\label{sec:second}

We then test if second-order effects are present in the residuals of the light 
curve. These effects may have several origins: orbital motion of the Earth 
around the Sun \citep{gould92}, which induces a deviation of the lens-source 
motion from rectilinear, orbital motion of the planet about the lens star, 
orbital motion of the source star if it is a binary, and possible additional 
objects (planets or stars) in the lens system. 

In fact we will limit our study to the first two effects, namely microlens 
parallax and planet orbital motion. A binary source in which the companion is 
either not lensed or too faint to contribute to the light curve can mimic the 
parallax effect due to the acceleration of the source, in a so-called 
``xallarap'' effect. This effect is described for instance in 
\citet{poindexter05} and tested in \citet{miyake12}. However, it implies in its
simplest form 5 additional parameters versus 2 each for parallax and planet 
orbital motion. Given the fact that our detection of second-order effects, 
although clear, remains relatively marginal, it would be difficult to trust 5
additional parameters constrained by the light curve residuals from a standard 
model. Moreover, binary stars generally have long periods, the effect of which
will remain undetected in the short lapse of time when our events are observed
(months versus years).

Testing for a third body, either from a circumbinary planet or a second planet
orbiting the lens star, is beyond the scope of this paper, and again would 
imply a large number of additional parameters to fit, without any guarantee to 
obtain meaningful results. In the short history of planetary microlensing, it 
has been evidenced only once \citep{gaudi08,bennett-ogle109}, and in that case
only because a feature in the light curve could not be fitted by a single 
planet model. There is no such feature in the present light curve. As the 
domain of possible configurations to test if we add a third body is vast, the 
effort expended would be incommensurate with the potential scientific return, 
at least until we get independent evidence that such a modeling effort is 
necessary: see Section~\ref{sec:HighRes}.

As discussed in Section~\ref{sec:overview}, the microlens parallax is 
characterized by a 2-dimensional vector $\bpi_{\rm E}$ whose magnitude is given
by Equation~(\ref{eqn:theta_e+pi_e}) and whose direction is that of the 
lens-source relative motion on the plane of the sky. Hence, there are two 
parameters $\bpi_{\rm E}= (\pi_{\rm E,N},\pi_{\rm E,E})$, the components of this 
vector in equatorial coordinates.

The planet orbital motion affects the light curve in two different ways. First,
it causes the binary axis to rotate or, equivalently, makes the source 
trajectory angle change in time \citep{dominik98,ioka99,albrow00,penny11}. 
Second, it causes the projected binary separation, and thus the magnification 
pattern, to change in time. Considering that the time scale of the lensing 
event is of order a month while the orbital period of typical microlensing 
planets is of order years, the rates of change in $\alpha$ and $s$ can be 
approximated as uniform during the event and thus the orbital effect is 
parameterized by
\begin{equation} \label{eqn:eq1}
\alpha(t) = \alpha(t_{\rm ref}) + 
\omega \times (t-t_{\rm ref}) 
\end{equation}
and
\begin{equation} \label{eqn:eq2}
s(t) = s(t_{\rm ref}) + \dot{s} \times (t-t_{\rm ref})
\end{equation}
where $\omega$ and $\dot{s}$ are the rates of change in the source trajectory 
angle and projected binary separation in units of ${\rm yr}^{-1}$, respectively,
and $t_{\rm ref}$ is a reference time. As explained in \citet{batista11}, the 
effect of planet orbital motion is similar to the microlensing parallax effect 
in the sense that the deviations caused by both effects are smooth and long 
lasting. Then, if the deviation caused by the planet orbital motion is modeled 
by the parallax effect alone, the measured parallax would differ from the 
correct value. Therefore, the orbital motion effect is important not only to 
constrain the orbital properties of the lens, but also to precisely constrain 
the lens parallax and thus the physical parameters of the lens system.  

\begin{deluxetable}{lcccc}
\tablecaption{Best-fit Model Parameters\label{tab:one}}
\tablewidth{0pt}
\tablehead{
\multicolumn{1}{c}{parameter} &
\multicolumn{4}{c}{model} \\
\multicolumn{1}{c}{} &
\multicolumn{1}{c}{standard} &
\multicolumn{1}{c}{parallax} &
\multicolumn{1}{c}{orbit+parallax ($u_0<0$)} &
\multicolumn{1}{c}{orbit+parallax ($u_0>0$)}
}
\startdata
$\chi^2$                   & 6420.051   & 6411.450   & 6369.345   & 6365.664   \\
$t_0$ (HJD')               & 5420.93685 & 5420.93702 & 5420.93773 & 5420.93915 \\
$u_0$                      & -0.003562  & -0.003540  & -0.003372  & +0.003404  \\
$t_{\rm E}$ (days)          & 42.55      & 42.94      & 46.57      & 46.92      \\
$s$                        & 1.12282    & 1.12338    & 1.12385    & 1.12279    \\
$q$                        & 0.0023808  & 0.0023694  & 0.0022075  & 0.0021809  \\
$\alpha$ (rad)             & 3.67605    & 3.67528    & 3.68023    & 2.60087    \\
$\rho_\star$                & 0.0006429  & 0.0006403  & 0.0005861  & 0.0005764  \\
$f_s$                      & 0.4111     & 0.4082     & 0.3757     & 0.3731     \\
$\sqrt{f_s}/\rho_\star$     & 997.3      & 997.9      & 1045.8     & 1059.8     \\
$\pi_{{\rm E},N}$            & --         & +0.37      & +0.27      & +0.77      \\
$\pi_{{\rm E},E}$            & --         & +0.012     & +0.02      & -0.11      \\
$\dot{s}$ (${\rm yr}^{-1}$) & --         & --         & +0.64      & +0.86      \\
$\omega$ (${\rm yr}^{-1}$)  & --         & --         & +0.06      & -1.28     \\
$T_{\rm orb}$ (yr)           & --         & --         & 14.25      & 4.59      \\
$t_{\rm ref}$ (HJD')         & --         & 5421       & 5421       & 5421      \\
\enddata
\tablecomments{
$\rm  HJD'=HJD-2450000$. $t_{\rm ref}$ is the reference time of the model, when 
the model reference frame moves at the same speed as the Earth and 
$\alpha(t) = \alpha$ and $s(t) = s$.}
\end{deluxetable}

For each tested model, we search for the solution of the best-fit parameters 
by minimizing $\chi^2$ in the parameter space. In the initial search for 
solutions, we divide the parameters into two categories. Parameters in the 
first category are held fixed during the fitting, and parameter space is 
searched with a grid. For the parameters in the second category, solutions are 
searched by using a downhill approach. We choose $s$, $q$, and $\alpha$ as the 
grid parameters because they are related to the light curve features in a 
complex way, where a small change in the parameter can result in dramatic 
changes in the light curve. On the other hand, the remaining parameters are 
more directly related to the features of the light curve and can thus be 
searched by using a downhill approach. For the downhill $\chi^2$ minimization, 
we use a Markov Chain Monte Carlo method. Brute-force search throughout the 
grid-parameter space is also needed to test the possibility of the existence of 
local minima that result in degenerate solutions. For the light curve of 
MOA~2010-BLG-477, we find that the other local $\chi^2$ minima have $\chi^2$ 
values much larger than the best fit by $\Delta\chi^2 \simgt 6000$ and are 
therefore not viable solutions. Once the space of the grid parameters around 
the solution is sufficiently narrowed down, we allow the grid parameters to 
vary in order to pin down the exact location of the solution and to estimate 
the uncertainties of the parameters.

Modeling was also done independently using the method of \citet{bennett-himag}, 
and this analysis reached the same conclusions. This independent analysis also 
uses a slightly different implementation of the planetary orbital motion 
parameters \citep{bennett-ogle109}. Equations~\ref{eqn:eq1} and \ref{eqn:eq2} 
describe the orbital velocities, which are the first order contribution of 
orbital motion. To second order, we have only one component of acceleration 
because this must be directed towards the host star. But, one additional 
parameter is also all that is needed to describe a circular orbit, so we can 
add the planetary orbital period, $T_{\rm orb}$ as a parameter and replace the 
constant velocities (in polar coordinates) of Equations~\ref{eqn:eq1} and 
\ref{eqn:eq2} with the projection into the plane of the sky of the circular 
orbit described by $s$, $\alpha$, $\dot s$, $\omega$, and $T_{\rm orb}$. Unlike 
the case of OGLE~2006-BLG-109Lb,c \citep{bennett-ogle109}, the value of 
$T_{\rm orb}$ does not have an influence on the light curve model $\chi^2$ 
values.

However, $T_{\rm orb}$ is still useful because it can be used to help constrain
the other orbital parameters to values consistent with a physical orbit. 
(This is an issue because it is quite possible to have $\dot s$ and $\omega$ 
values that are not consistent with a bound orbit, and this is a simple way to 
ensure that this is not the case). If we assume that $\theta_*$ is known, then 
we can calculate the lens system mass, $M = \theta_{\rm E}/(\kappa\pi_{\rm E})$, 
which follows from Equation~\ref{eqn:theta_e+pi_e}. This also allows us to 
determine $\pi_{\rm rel}$, but we cannot determine the lens and source 
distances, $D_L$ and $D_S$, separately. However, we can use the $T_{\rm orb}$ 
value to determine the orbital semi-major axis, under the assumption that the 
orbit is circular. This allows us to determine the Einstein radius, 
$R_{\rm E} = D_L\theta_{\rm E}$ in physical units, and since 
$R_{\rm E}^2 = 4GMD_L^2 D_S \pi_{\rm rel}/(c^2{\rm AU})$, we have a second 
relation between $D_S$ and $D_L$. But, since the source star is very likely to 
be in the Galactic Bulge, we also have approximate knowledge of $D_S$. 
Therefore, we can apply a constraint on the value of $D_S$ implied by the light 
curve parameters, $D_S = 8.0 \pm 1.2\,$kpc.

For this event, $T_{\rm orb}$ is not really constrained by the light curve 
measurements, so this constraint serves to force $T_{\rm orb}$ toward a value 
consistent with a circular orbit for a Bulge source. This constraint on the 
source distance also serves to enforce a constraint on the orbital velocity 
parameters, $\dot s$ and $\omega$. They must also be consistent with a circular 
orbit for a Bulge source. Parameters that satisfy this constraint are also 
consistent with most orbits with moderate eccentricity, $\epsilon \simlt 0.5$.
But, these parameters are not consistent with orbits with the highest possible 
transverse velocities. In fact, the light curve measurements marginally 
favor implausibly large $\dot s$ and $\omega$ values corresponding to orbits 
that are either unbound or just barely bound. These barely bound orbits 
with large $\dot s$ and $\omega$ values have high eccentricities that just 
happen to have been observed with motion in the plane of the sky during the 
brief time near periapsis. The best fit model that is consistent with a bound 
orbit is such a model, which has an orbit with an extremely low {\it a priori} 
probability. This low {\it a priori} probability makes such a model much more 
unlikely than the best fit model with the circular orbit constraint, so we 
report the best fit model with the circular orbit constraint as the ``best fit 
model" in Table~\ref{tab:one}.

However, the unlikely models with values of $\dot s$ and $\omega$ do tend to
have better light curve $\chi^2$ values, and they also cover a large 
volume of parameter space. So, they should not be ignored in our consideration 
in the range of possible physical parameters for the MOA~2010-BLG-477L 
planetary system. Therefore, we do not enforce this source distance constraint 
in the MCMC runs that we use to estimate the distribution of likely physical 
parameters for this system. Instead, we apply a more general constraint on the 
orbital and Galactic parameters of the lens system as discussed in 
Appendix~\ref{app:jacobi}.

\section{Results}\label{sec:results}

\begin{figure}[th]
\epsscale{1.0}
\plotone{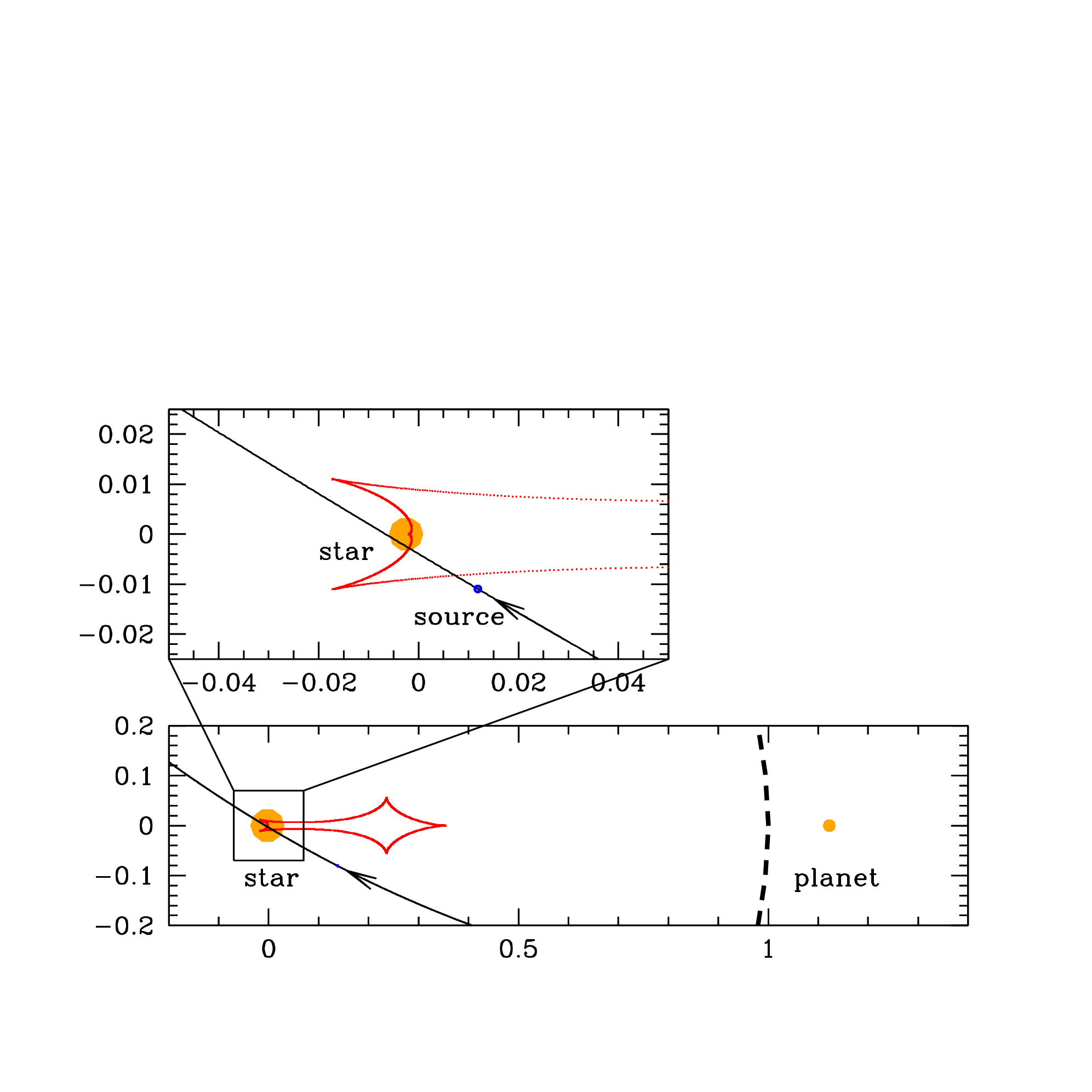}
\caption{\label{fig:geom}
Geometry of the lens system responsible for the microlensing event 
MOA~2010-BLG-477. In the lower panel, the filled dots represent the 
locations of the host star and planet. The closed figure composed 
of concave curves represents the caustic. The curve with an arrow 
represents the source trajectory with respect to the caustic. 
The big dashed circle centered at the host star represents the Einstein 
ring.  The upper panel shows the enlargement of the region enclosed 
by a small box in the lower panel. The small filled circle represents 
the source star where its size is scaled by the caustic size.}
\end{figure}

In Table~\ref{tab:one}, we present the results of modeling along with 
the best-fit parameters for the 3 tested models.  The best-fit light curve 
is presented in Figure~\ref{fig:lcurve}. In Figure~\ref{fig:geom}, we also 
present the geometry of the lens system. It is found that the perturbation 
near the peak of the light curve was caused by the source crossings and 
approaches of the caustic produced by the binary system with a low-mass 
companion. The measured mass ratio between the lens components is 
$q=(2.181\pm0.004)\times 10^{-3}$ and thus the companion is very likely to 
be a planet.  The measured projected separation between the lens components 
is $s=1.1228\pm0.0006$, which is close to the Einstein radius.  As a result, 
the caustic is resonant, implying that the caustic forms a single 
closed curve composed of 6 cusps.  The perturbations at $\rm HJD'\sim 5420.4$ 
and $5420.9$ are produced by the source crossings of one of the star-side 
tips of the caustic. The bumps at $\rm HJD'\sim 5421.0$ and $5422.4$ are caused
by the source's approach close to the weak and strong cusps on the side of the 
host star, respectively.  

\subsection{Constraints from Measurement of $\theta_{\rm E}$}\label{sec:source}

\begin{figure}[th]
\epsscale{0.8}
\plotone{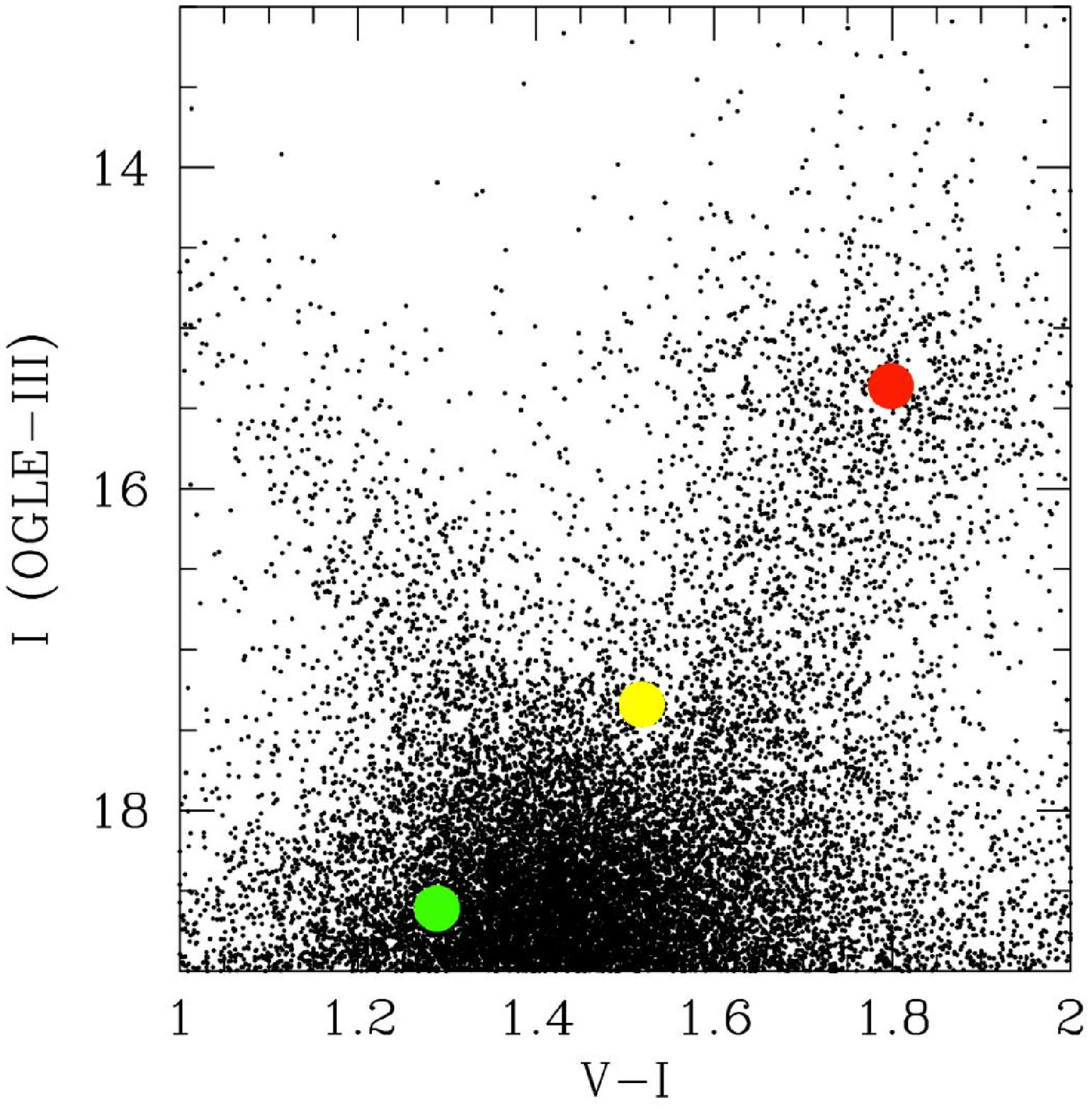}
\caption{\label{fig:cmd}
OGLE-III 8x8 arcmin calibrated color-magnitude diagram (CMD) in $I$- and 
$V$-band. The red, yellow and green solid circles mark the centroid of the red 
giant clump, the blend, and the source position, respectively.}
\end{figure}

The most important constraints on the lens mass and distance come from the 
measurement of $\theta_{\rm E}$, which can be rewritten from 
Equation~(\ref{eqn:rhodef}) as
\begin{equation}
\theta_{\rm E} = \frac{\theta_\star}{\rho_\star} = \frac{\sqrt{F_I/(\pi S_I)}}{\rho_\star}
\label{eqn:thetae_dependence}
\end{equation}
where $F_I$ is the intrinsic (dereddened) flux of the source and $S_I$ is its 
intrinsic $I$-band surface brightness. A key point is that the surface 
brightness does not depend at all on the microlens model (just on the source 
color and/or spectrum). Hence, $\theta_{\rm E}$ depends on the model only 
through the parameter combination $\sqrt{f_s}/\rho_\star$, where $f_s$ is the 
instrumental source flux ($f_s=1$ corresponds to magnitude 18). It is clear 
from Table~\ref{tab:one} that this parameter combination varies very little. 
The biggest uncertainties are therefore in measuring the surface brightness and
measuring the offset between the instrumental flux $f_s$ and the dereddened 
source flux $F_I$. Traditionally, these are measured simultaneously by 
determining the offsets of the color and magnitude, respectively, of the source
from the clump on an instrumental CMD \citep{yoo04}. An instrumental CMD from
CTIO is shown in Figure~\ref{fig:cmd}. From it, we can read instrumental 
magnitude in $I$ of 19.07. The instrumental $V-I$ color is measured more
accurately from a regression of $V$ flux vs. $I$ flux, and gives $-0.35$.

Comparing to the instrumental clump position, we find $(V-I)_0=0.55\pm 0.05$ 
and $I_0=17.61\pm 0.15$, assuming $[(V-I),I]_{0,\rm clump} = (1.06,14.42)$. The 
color error is determined empirically from a sample of microlensed dwarfs with 
spectra \citep{bensby11}, while the magnitude error comes primarily from the 
error in fitting the clump centroid and the assumed Galactocentric distance of 
8~kpc (both about 0.1 mag). These lead to an estimate 
$\theta_\star = 0.79\pm 0.06$.

We can compare the dereddened color estimate to the color deduced from the 
high-resolution spectrum reported in Section~\ref{sec:obs}, from which we 
measured an effective temperature of $T=5950\pm 150\,$K, which corresponds to 
$(V-I)_0=0.65\pm 0.04$. These determinations are marginally consistent. We 
adopt the first to maintain the general practice of microlensing papers, but 
note that if we adopted the mean of the two estimates, the inferred value of 
$\theta_\star$ (and so $\theta_{\rm E}$) would rise by 2\%.

Finally, we evaluate
\begin{equation}
\theta_{\rm E} = 1.38\pm 0.11\,{\rm mas}
\label{eqn:thetae_eval}
\end{equation}
This value is unusually large and implies that the lens must be very massive or
very close. Specifically
\begin{equation}
\frac{M}{M_\odot} \frac{\pi_{\rm rel}}{{\rm mas}} =
\frac{\theta_{\rm E}^2}{\kappa\,M_\odot\,{\rm mas}} = 
0.233\pm 0.036
\label{eqn:thetae2}
\end{equation}

{\subsection{Constraints from Lens Flux Limits}
\label{sec:flux}}

The model gives a measurement of the light coming not only from the source, but
also from any other stars in the same point spread function, generally called 
blended light. This light may come from the lens itself, a companion to the 
lens, a companion to the source, or any unrelated star on the same line of 
sight, but not participating to the amplification process. 

\begin{figure}[th]
\epsscale{1.0}
\plotone{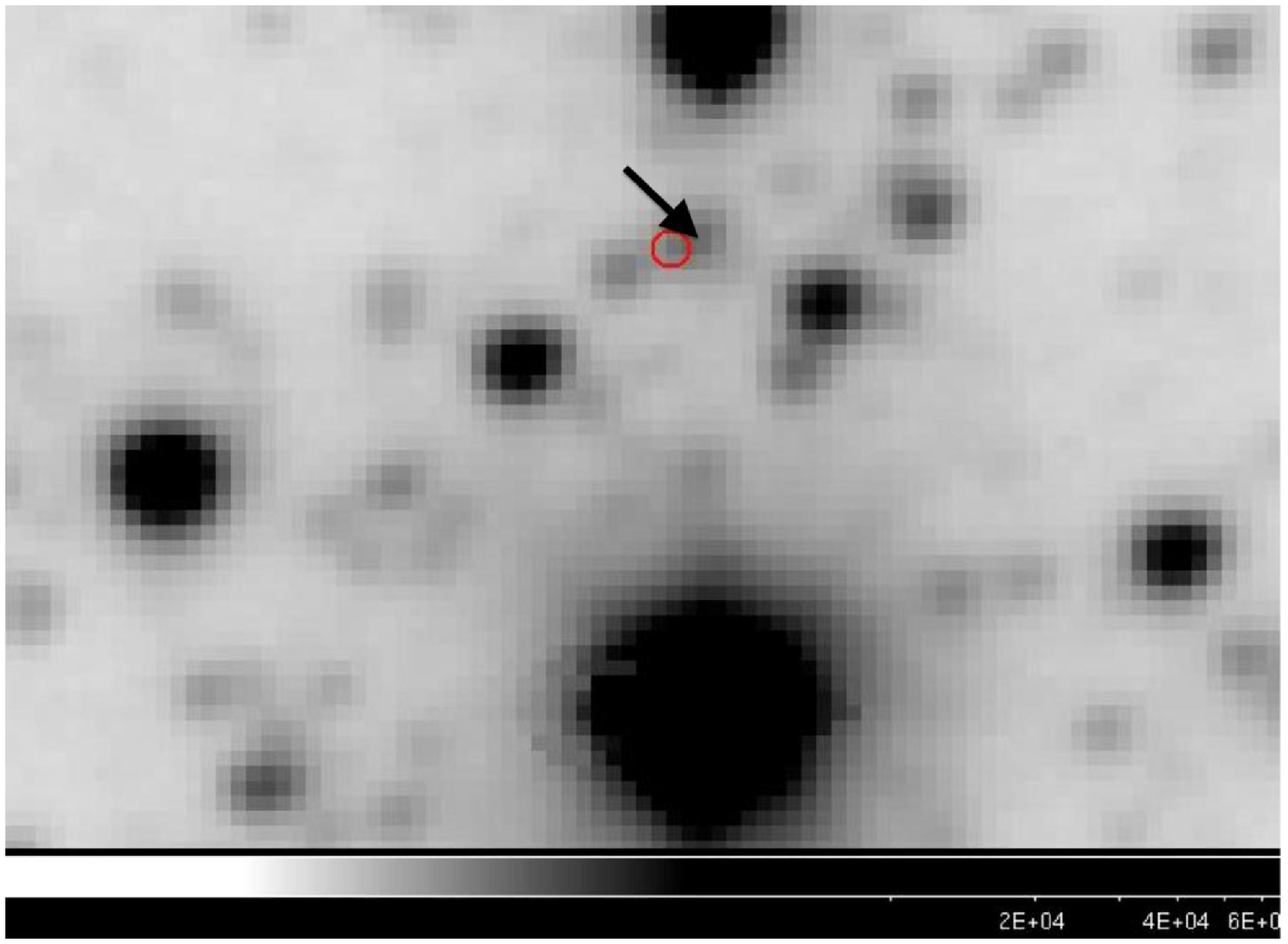}
\caption{\label{fig:ogle}
OGLE-III reference image of the BLG 176.8 field, in the region of our 
microlensed target. North is up and East to the left side. One pixel is 
$0.26 \arcsec$. The red circle marks the position of the source (and the lens) 
from DIA analysis of CTIO images, in between two OGLE stars: the brighter of
them (pointed at by a black arrow), at $0.37 \arcsec$ from the source, 
corresponds in magnitude to the blended light measured by the model. The source 
is too faint to be detected in this crowded image.}
\end{figure}

The OGLE-III image of this field, displayed in Figure~\ref{fig:ogle}, shows two
stars at the target position. Their $I$-band magnitudes in the OGLE-III
photometric catalogue is $I=17.446\pm0.052$ for the brighter star, numbered
119416, and $I=17.807\pm0.059$ for the fainter one with number 119534. They 
are separated by $1.26\arcsec$, enough to be separated by PSF photometry at
good sites such as CTIO or Las Campanas (OGLE) in Chile. A difference image
analysis (DIA) made on CTIO images shows that the microlensed star does not
correspond to the position of either star (red circle in Figure~\ref{fig:ogle}),
but is displaced by 1 CTIO pixel ($0.37\arcsec$) from the brighter star.

Now, the blended light in $I$-band as measured from DoPhot CTIO photometry by 
the model is $I_{\rm b}=17.443\pm0.031$, and this corresponds precisely to the 
flux from the brighter star among the two OGLE stars, which is not separated 
from the microlensed target at the scale of the CTIO seeing (typically 
$1\arcsec$). Knowing that the blended flux comes from an unrelated blended 
star, the light from the lens must be smaller. We can rigorously conclude that 
the lens has less than half the light in the observed blend. Otherwise, the 
lens and blend would be separated by at least $2\times 0.37''=0.74''$ in the 
OGLE image, and so would have been at least marginally resolved.

An additional argument showing that the lens is faint enough to remain 
undetected comes from its large relative proper motion ($10.3\pm0.8$~mas/yr). 
DIA analysis of OGLE-III good seeing images separated by 3.3 years shows no 
residual at the target position, which proves that no detected star has moved 
during this period.

Combining this limit with Equation~(\ref{eqn:thetae2}) yields strong 
constraints on the lens. The lens must be closer than the source, and so be at 
or closer than the Galactocentric distance, and suffer the same or less 
extinction. These imply $I_{\rm L,0}=I_{\rm b}-2.5\log(0.5)-A_{\rm I}>17.0$, and 
so $M_I>2.5$, which corresponds to $M<1.4\ M_\odot$ 
\citep{straizys81,bessell88}. Then, even the $2\,\sigma$ limit from 
Equation~(\ref{eqn:thetae2}) implies $\pi_{\rm rel} > 0.10$~mas and so 
$D_L <4.4\,$kpc. But $I_{\rm L,0}>17.0$ then implies $M_I>3.8$, which 
corresponds to $M<1.0\,M_\odot$. Cycling through this argument one more time 
yields $D_L <3.7\,$kpc, $M_I<4.2$, $M<1\ M_\odot$. If the lens were in front of 
some of the dust, this argument would become still stronger. However, at the 
relatively high latitude of this field ($b=-5^\circ$), most of the dust 
probably lies in front of 3~kpc, and in any case, there is no basis for 
adopting a more optimistic assumption about the dust.

\subsection{Constraints from the Microlens Parallax $\pi_{\rm E}$}

\begin{figure}[th]
\epsscale{0.75}
\plotone{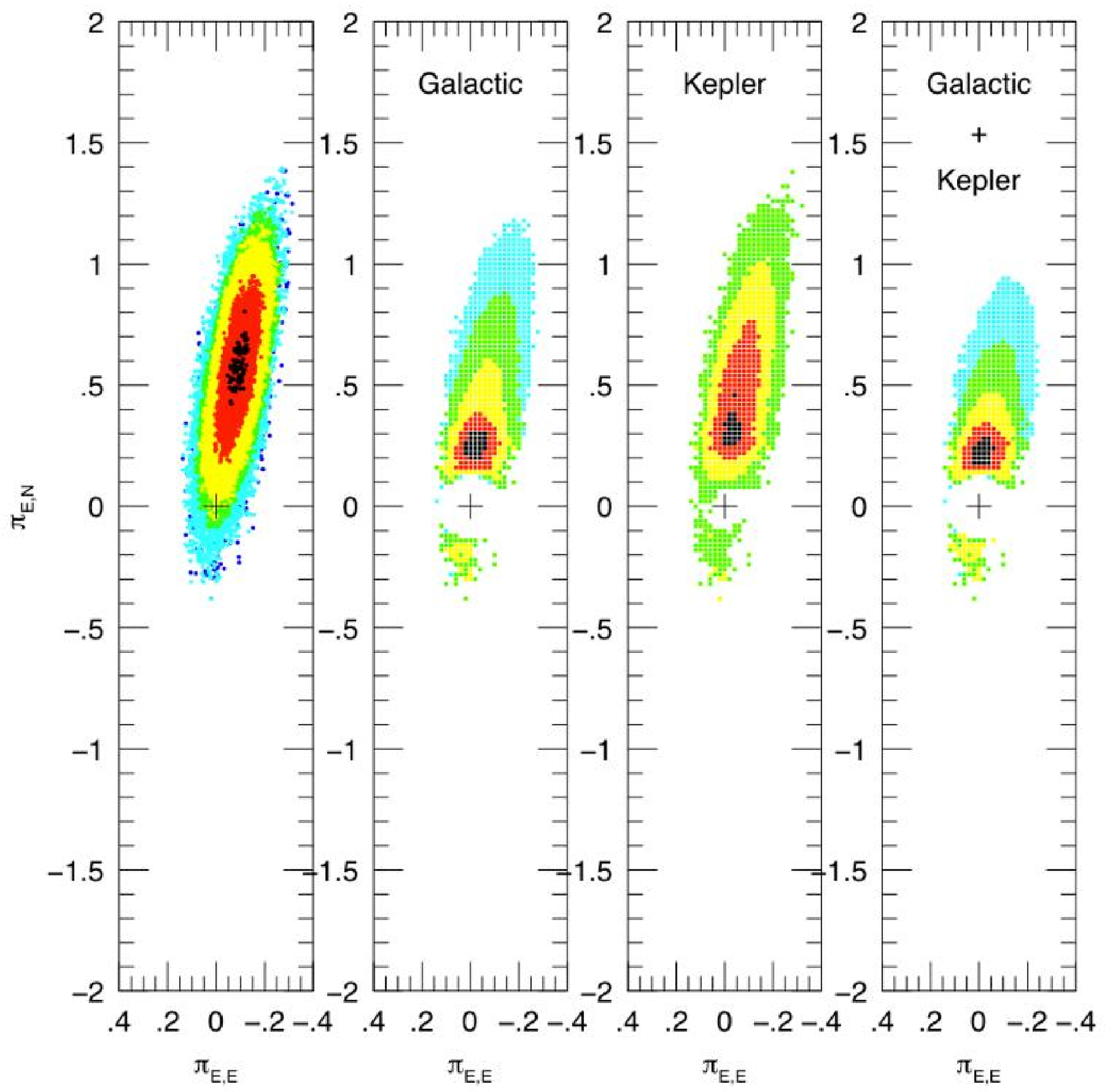}
\caption{\label{fig:bayes1}
Parallax vector $(\pi_{\rm E,N},\pi_{\rm E,E})$ for the $u_0>0$ solution, from 
Markov Chain Monte Carlo fit to MOA-2010-BLG-477. Left panel (a) displays 
individual chain points, color-coded by $\Delta\chi^2$ (1, 4, 9, 16, 25, 36) 
for black, red, yellow, green, cyan, blue. Second panel (b) shows effect of 
weighting by the Galactic model, geometric Jacobian, and flux constraint. Lower 
parallaxes (and so higher masses) are heavily favored, except that too low 
parallaxes (corresponding to $M>1.0\,M_\odot$) are ruled out by the flux 
constraint. Third panel shows the effect of weighting using the Jacobian due to 
orbital motion parameters, which by itself disfavors the heaviest masses 
because these tend to imply unphysical orbits, but favors moderately high 
masses (see text). The final panel shows the effect of combining Galactic and 
Kepler priors.}
\end{figure}

Table~\ref{tab:one} shows that parallax alone improves the fit by
${\Delta}{\chi^2}=8.6$, while including both parallax and lens orbital motion 
improves it by ${\Delta}{\chi^2}=54.4$. However, while {\it detection} of these
effects is therefore unambiguous, we cannot fully disentangle one from the 
other. For each effect, one of its two parameters is well determined while the 
other is highly degenerate with one parameter from the other effect. As first 
discovered by \citet{batista11} and further analyzed by \citet{skowron11}, weak
detections of parallax and orbital motion lead to a strong degeneracy between 
$\pi_{\rm E,\perp}$ (the component of $\bpi_{\rm E}$ perpendicular to the 
projected position of the Sun) and $\omega$ (the component of orbital motion 
perpendicular to planet-star axis), while $\pi_{\rm E,\parallel}$ and $\dot s$ 
are well constrained. The impact of this is illustrated in Panel (a) of 
Figure~\ref{fig:bayes1}, which shows $\chi^2$ values by color as a function of
$(\pi_{\rm E,N},\pi_{\rm E,E})$.  

First, it is clear that the parallax vector is almost completely degenerate 
along a line that is $8.4^\circ$ west of north, within $1^\circ$ of the 
predicted orientation of $\pi_{E,\perp}$ ($7.7^\circ$). Second, the light 
curve excludes  $\pi_{\rm E}>1.3$ at $3\,\sigma$. As we have
\begin{equation}
\frac{M}{M_\odot} \times \pi_{\rm E} =
\frac{\theta_{\rm E}}{\kappa\,M_\odot\,{\rm mas}} = 
0.169\pm 0.013
\label{eqn:thetae}
\end{equation}
this corresponds to $M>0.13\,M_\odot$. Thus, combining constraints from this
section and from Section~\ref{sec:flux}, and using 
Equation~(\ref{eqn:thetae2}), we have
\begin{equation}
0.13\,M_\odot<M<1.0\,M_\odot;
0.5\,{\rm kpc}<D_L<2.8\,{\rm kpc}
\label{eqn:massdistlims}
\end{equation}
Note that, because the parallax contours pass through the origin, parallax 
provides no additional constraint at low $\pi_{\rm E}$, i.e., at high mass.

\begin{figure}[th]
\epsscale{0.75}
\plotone{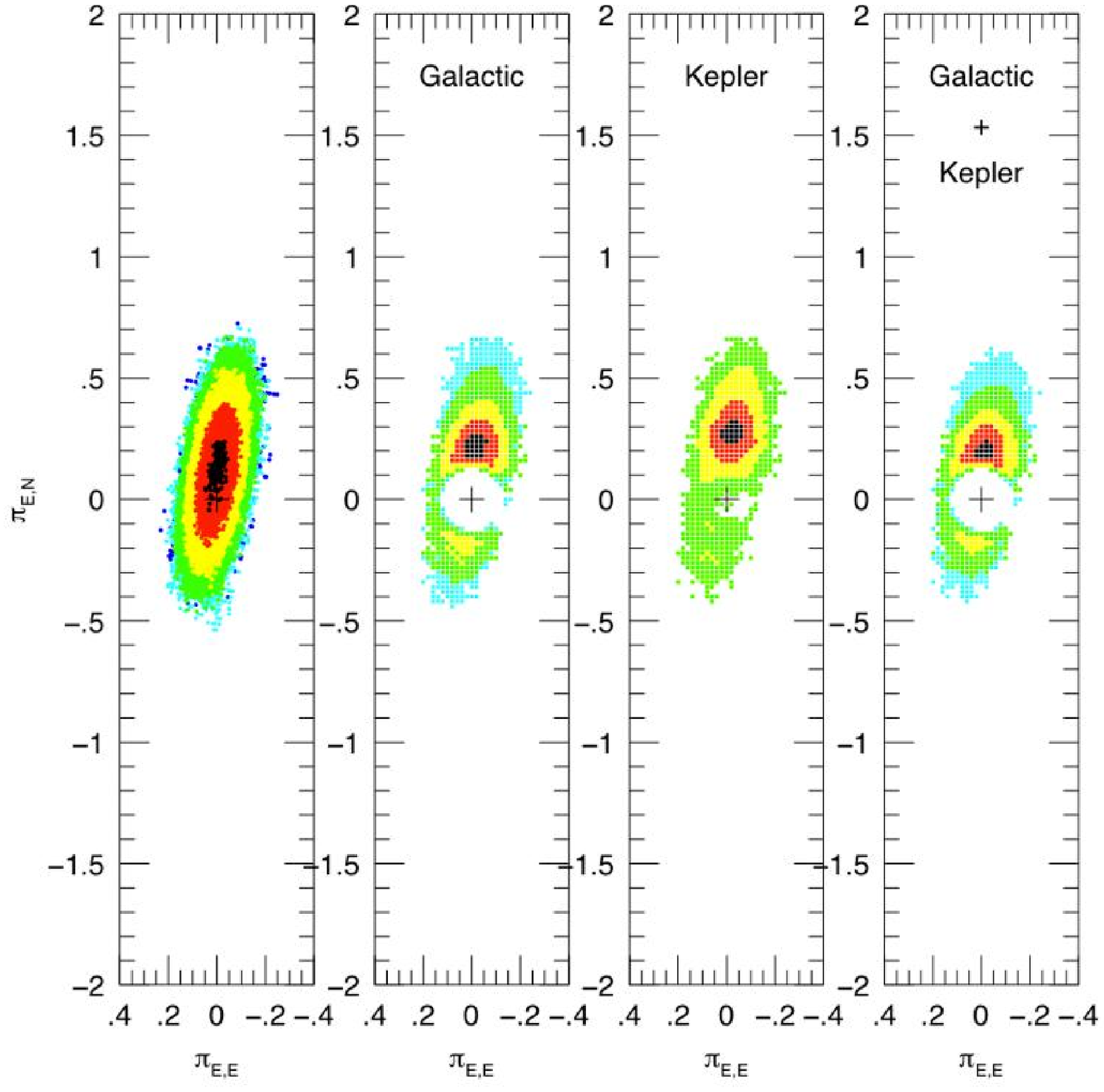}
\caption{\label{fig:bayes2}
Parallax vector $(\pi_{\rm E,N},\pi_{\rm E,E})$ for the $u_0<0$ solution, from 
Markov Chain Monte Carlo fit to MOA-2010-BLG-477. See description of
Figure~\ref{fig:bayes1} for details.}
\end{figure}

A similar diagram is given for the slightly disfavored $u_0<0$ solution in
Figure~\ref{fig:bayes2}.

\subsection{Post-Bayesian Analysis}

Equation~(\ref{eqn:massdistlims}) defines the limits of what can be said 
rigorously about the lens mass and distance based on current data. As we 
discuss in Section~\ref{sec:HighRes}, the lens could almost certainly be 
detected by high-resolution imaging, which would probably completely resolve 
the uncertainty in Equation~(\ref{eqn:massdistlims}). In the meantime, we can 
perform a Bayesian analysis based on a Galactic model and constraints from a 
Keplerian orbit. Figure~\ref{fig:bayes1} shows that each of these constraints 
separately favors small values of $\pi_{\rm E}$ and hence, relatively large 
masses and distances (within the limits set by 
Equation~[\ref{eqn:massdistlims}]), and so tend to reinforce each other. 
Panel~(a) shows the raw results of the MCMC.  Panel~(b) shows the same chain,
post-weighted by the Bayesian prior due to the Galactic model and the flux 
constraint.  The latter, which implies $M<1.0\,M_\odot$ 
(Section~\ref{sec:flux}) is responsible for the circular ``holes'' at the
centers of panels (b) and (d). As discussed in detail by \citet{batista11} this 
includes terms not only reflecting the density of lenses along the sight and 
the expected distribution of proper motions, but also a Jacobian transforming 
from the ``natural microlensing variables'' to the physical system of the 
Galaxy. It is this last term that actually dominates, particularly in the 
``near field'' ($D_L<2.8\,$kpc) permitted by the constraints, and is roughly 
$\sim D_L^5$ (\citet{batista11}, Equation~[18]). This weighting is so severe 
that one must ask whether the result of the weighting is plausibly compatible 
with the raw $\chi^2$ from the light curve. In this case, the solutions favored 
by the Bayesian post-weighting are compatible with the raw $\chi^2$ minimum at 
better than $2\,\sigma$, so there is no major conflict.

Panel~(c) shows the results of post-weighting the chain only by the orbital 
Jacobian. This is described in detail by \citet{skowron11} for the case of 
chains with complete orbital solutions. These contain two orbital parameters 
(called $\gamma_z$ and $s_z$) in addition to the two first-order orbital 
parameters considered here ($\dot s$ and $\omega$). These higher order 
parameters would be completely unconstrained in the present problem, so we 
simply resample the chain with a uniform integration over these two parameters.
Panel~(c) clearly also favors more massive, more distant lenses (although not 
the most massive), but it is not immediately obvious why. 
Appendix~\ref{app:jacobi} details the reasons behind this result. Finally, we 
note that, as expected, the combined effect of these two priors shown in 
Panel~(d) is stronger than either separately.

Why do the Galactic model and Kepler constraints each favor more distant 
(and more massive lenses) than the light curve alone (see 
Fig.~\ref{fig:bayes1})? The Galactic model constraint is virtually guaranteed 
to favor more distant lenses because, as mentioned above, most of the weighting 
is simply due to a coordinate transformation from microlensing to physical 
coordinates, and there is more phase space at larger distances. Since the 
distance errors are fairly large, this effect will be relatively strong. By 
contrast, the Kepler constraint could have just as easily favored more-distant 
as less-distant lenses. The most likely explanation is then, simply, that the 
light curve prediction of the distance is too close by 1.7 sigma, and the 
constrained value (panel d) is a better estimate. This conjecture is testable 
by future observations, as described in the next section.

\section{Future High Resolution Observations}\label{sec:HighRes}

Follow-up observations are important to check the predictions of our 
models. In the microlensing field, the idea of doing follow-up observations
preceded the detection of planets, going back at least to 1998 in the case of 
MACHO-LMC-5, with the corresponding HST observations published in 
\citet{alcock01}. Then, it was successfully applied to derive more accurate 
parameters of the planetary systems, using HST or adaptive optics systems at 
Keck or VLT, for instance in 
\citet{bennett06,dong09,janczak10,batista11,kubas12}. In the future, it could 
be used to study planets in the habitable zones of nearby dwarf stars, as 
suggested by \citet{distefano08,distefano12}.

While the Bayesian analysis strongly favors a lens close to $1\,M_\odot$,
i.e, the upper limit permitted by the measurements, there is no airtight 
evidence against lower-mass lenses. Fortunately, this question can almost 
certainly be resolved by high-resolution imaging. The $VIH$ source magnitudes 
are $V_s=19.98\pm0.03$, $I_s=18.71\pm0.03$, and $H_s=17.35\pm0.03$. These come 
from CTIO $H$-band measurements, taken simultaneously with $I$ and $V$-band
observations. A linear regression between both fluxes gives a very accurate 
instrumental color of $(I-H)_{\rm CTIO}=1.707\pm0.002$. After converting $H$ to 
the 2MASS photometric system and $I$ to the OGLE-III system by use of common 
stars in the CTIO field, the color becomes $I-H=1.36\pm0.03$. Subtracting this 
color from the $I$ magnitude of the source in the OGLE-III system ($18.71$) 
returns the above value for $H$.

As we now show, the lens brightness must be similar to the source brightness. 
As can be judged from Figure~\ref{fig:ogle}, the field is relatively sparse,
at a Galactic latitude of $b = -5.095^\circ$, so these two stars (for the moment
superposed) will very likely be the only two stars in their immediate 
high-resolution neighborhood.

Let us consider three examples consistent with Equation~(\ref{eqn:thetae2}),
$(M/M_\odot,D_L/{\rm kpc})=(0.1,0.4)$, $(0.5,1.7)$, $(1.0,2.8)$.
These would have lens absolute magnitudes $M_H=(9.6,6.0,3.4)$ according to
\citet{kroupa97} and so $H_{\rm L,0}=(17.6,17.1,15.6)$. Of course, the 
extinction would be different at these different distances, but the entire 
column to the source is only $A_H\sim 0.3$. Therefore, a 0.5~$M_\odot$ lens 
star will be as bright as the source, and even a lens star at the bottom of the 
main sequence will produce an easily detectable amount of light ($0.5$~mag) 
over the expected source magnitude.

Since all $M/D_L$ combinations produce similar $H$ magnitudes, such a 
measurement would, by itself, have little predictive power. But these various 
scenarios would yield substantially different $J-H$ colors, which would add 
discriminatory power. Time has been allocated on various large telescopes to
observe the field of this event, detect and measure the light coming from the
source and lens stars. 

In addition, because of the relatively large proper motion,
$\mu=10.3\pm0.8$~mas/yr the lens and source could be separately resolved 
within about 5 years. This would then yield the angle of proper motion 
($=\tan^{-1}(\pi_{\rm E,E}/\pi_{\rm E,N})$) and so (from Figure~\ref{fig:bayes1}) 
the amplitude of $\pi_{\rm E}$ \citep{ghosh04}.

We can also consider follow-up observations with the Hubble Space Telescope,
which will be able to detect the lens-source relative proper motion as early as
2012 \citep{bennett07}. The implied absolute magnitudes for the three examples
given above ($M = 0.1$, 0.5, and $1.0\,M_\odot$) are $M_I =(12.3,7.9,4.1)$, and 
the implied extinction free magnitudes are $I_{\rm L,0} = (20.3,19.1,16.3)$. 
Since the extinction in the foreground of the lens is $A_I < 1.0$, this implies 
that the host star should be detectable with at least 9\% of the $I$-band flux 
of the source star over the full range of main sequence host star masses.

Finally, the ESA satellite GAIA to be launched in 2013 will image the Galactic 
Bulge down to $V \sim 20$~mag, so it may detect the lensing object and will 
certainly measure its proper motion if it does.

In conclusion, these additional observations should reveal whether our 
choice of second-order effects (microlensing parallax and planet orbital 
motion) correspond to the reality. If we find contradictory results to our 
predictions, then it will be a strong argument to conduct the modeling of 
additional effects mentioned in Section~\ref{sec:second}, such as xallarap 
effect from a binary source, or involvement of a third body.

\section{Final results, conclusions and perspectives}

In order to select the best among the three competing models (standard,
parallax only, orbital motion and parallax), a simple comparison of $\chi^2$
values is not enough, because more refined models use more parameters. Taking
into account these additional parameters by normalizing the $\chi^2$ estimate
by the number of degree of freedom is not the proper way to select the best
model. A vast literature exists about model selection, and an application of
different criteria to astrophysics is described in \citet{liddle07}. A simple
way to take care of the larger number of parameters is to use the Akaike 
Information Criterion (AIC) \citep{akaike74}, which introduces a penalty to the
$\chi^2$ by adding twice the number of additional parameters. Different 
criteria, such as the Bayesian Information Criterion (BIC) \citep{schwarz78} or
the Deviation Information Criterion, introduced by \citet{sbcl02}, can also be 
used. Formulas are given below, where $\Delta k$ is the number of additional 
parameters, $N$ the number of data points, and $\chi^2 (\overline{\theta})$ the 
$\chi^2$ of the average parameter set $\overline{\theta}$.

\begin{eqnarray} \label{eqn:eq11}
\Delta {\rm AIC} & = & \Delta \chi^2 + 2 \  \Delta k \nonumber \\
\Delta {\rm BIC} & = & \Delta \chi^2 + \Delta k \  \log N \nonumber \\
\Delta {\rm DIC} & = & \Delta (2 \  \overline{\chi^2} - \chi^2 (\overline{\theta}))
\end{eqnarray}

As there are 7 parameters in the standard model, 9 in the parallax only model 
and 11 in the orbital motion and parallax model, we see that our observed 
difference in $\chi^2$ of 8.6 in the parallax only model is only marginally 
significant according to the AIC (the expected difference is 4), while the 
observed difference of 54.4 (for the $u_0>0$) or 50.7 (for the $u_0<0$) is 
clearly an improvement of the orbital motion and parallax model over the 
standard one (the expected difference is 8). Similar results are obtained using
DIC, while the difference between models returned by BIC is less significant.

However, it is important to note that $\Delta k$ strictly corresponds to the 
number of additional parameters only in the case of a linear regression 
problem. Here, we clearly have non-linear fits, so we should compute an
``effective'' number of parameters, which is difficult to estimate. The above
conclusion should therefore not be taken as a quantitative one.

\begin{figure}[th]
\epsscale{1.0}
\plottwo{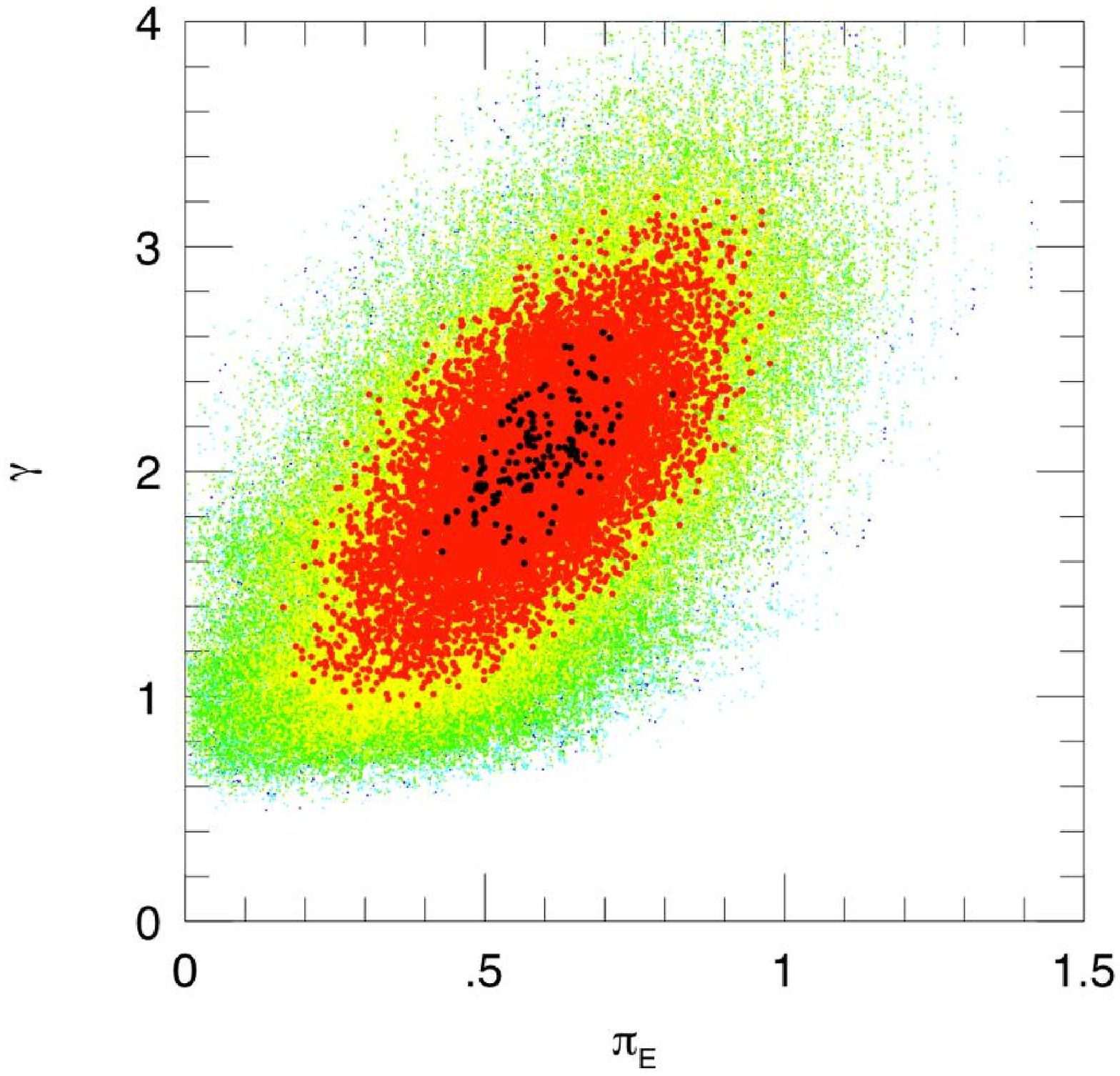}{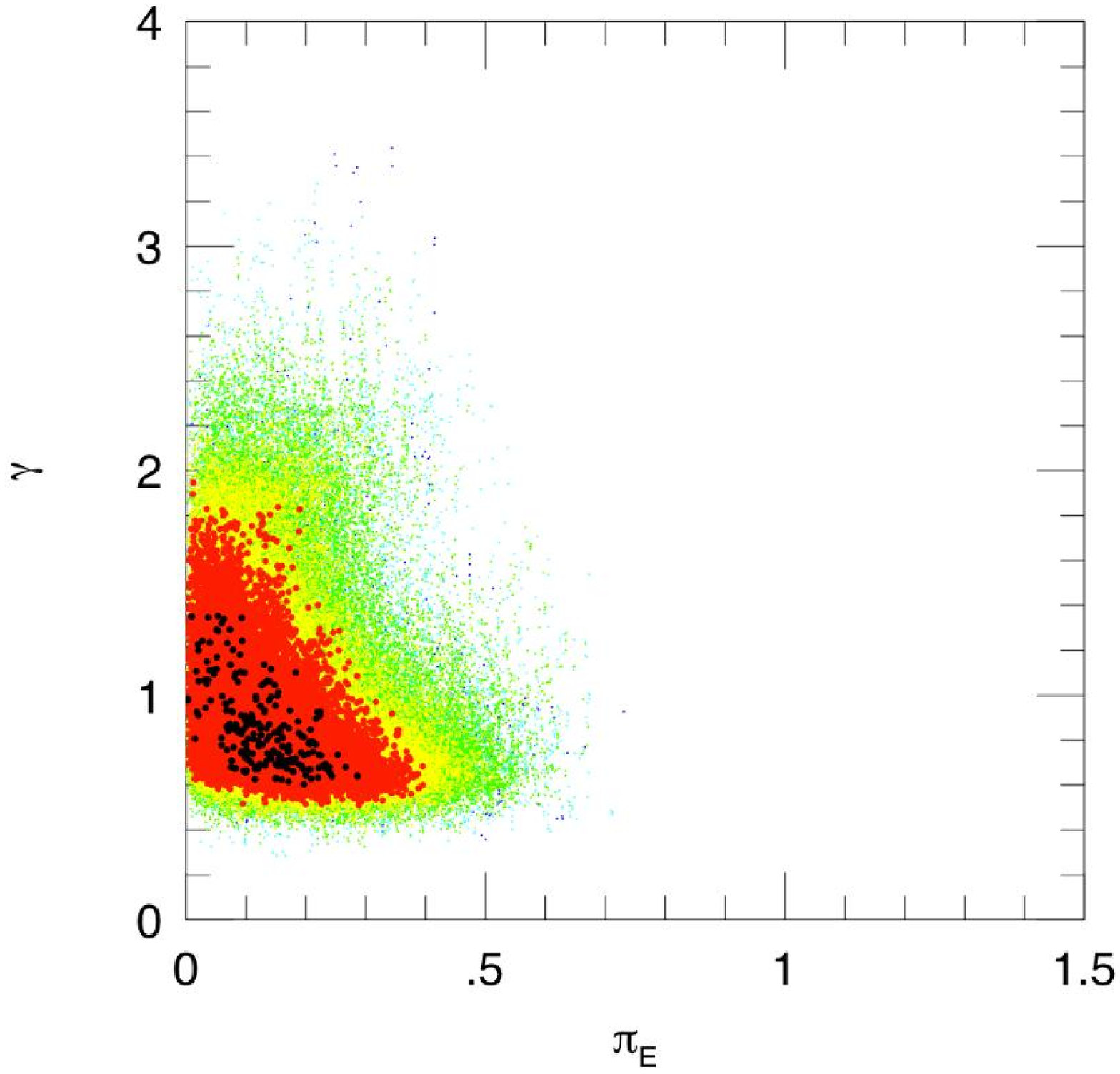}
\caption{\label{fig:secorder}
Variation of the orbital motion effect vs. the parallax effect for the orbital 
motion and parallax solutions. The left side corresponds to the $u_0>0$ 
solution, while the right side is for the $u_0<0$ solution.}
\end{figure}

A confirmation of the detection of second-order effects comes from the fact that
in the orbital motion and parallax model, the degeneracy between $u_0<0$ and
$u_0>0$ models is clearly broken. It is instructive to plot both second-order 
effects vs each other, separately for the $u_0<0$ and $u_0>0$ solutions. This is
done in Figure~\ref{fig:secorder}, where the orbital motion $\gamma$ is plotted 
vs. the parallax effect $\pi_{\rm E}$. 

If we remember that the ratio of the projected kinetic to potential energy
must be smaller than 1 to get bound orbits, and that this ratio is proportional
to $\gamma^2$, where the proportionality constant depends on $\pi_{\rm E}$ as
given by Equation~\ref{eqn:gamma0}, it is easy to interpret these diagrams. When
$\pi_{\rm E}$ increases, the proportionality factor decreases, so that if 
$\gamma$ remains small enough, the bound orbit condition is respected. In the
$u_0>0$ diagram (left side), small values of $\pi_{\rm E}$ (say below 0.17) have 
a proportionality factor larger than 1. As $\gamma$ is nearly 1 for these 
solutions, they are ruled out, and the light curve confirms it, as few small
$\chi^2$ solutions (black and red points) lie there. Larger $\pi_{\rm E}$ are 
also excluded, as they correspond to large $\gamma$ values, although the light 
curve would favor such solutions. The only surviving region in this diagram is 
around $\pi_{\rm E}\sim0.35$, corresponding to lens masses of half a solar mass, 
where the proportionality factor is about 0.4 and $\gamma$ slightly exceeds 1.

In the $u_0<0$ diagram (right side), although this is slightly disfavored by 
the light curve ($\Delta \chi^2=6.7$ for the best chain without the circular 
orbit constraint), there is a region where $\gamma$ is about constant at 0.7 
for $\pi_{\rm E}$ varying from 0 to 0.4. The bound solutions correspond to the 
larger values of the $\pi_{\rm E}$ domain (smaller proportionality factor), and 
they therefore agree with the range found in the $u_0>0$ diagram.

We therefore conclude that both solutions agree, and give bound orbits when the
lens mass is about half solar, corresponding to a lens distance of about 
1.6~kpc.

If we now move to the post-bayesian analysis, we see that this solution favored 
by the light curve has some tension with the Galactic model constraint, because
nearby lenses are rarer than more distant ones. But if we move to more distant 
lenses, we get many chains with unbound orbits or high eccentricities. By the
way, the $u_0>0$ solution, which has a lower $\chi^2$ than the alternate 
$u_0<0$ solution, is also the one where more chains correspond to unbound 
orbits.

There is therefore a tension between Galactic and Keplerian priors, and the 
issue will only be solved photometrically, by measuring the light coming from 
the source and the lens. This will be the subject of a forthcoming article 
about this event.

We conclude by giving the 1-D distributions of lens mass, lens distance, and 
planet orbit semi-major axis. The mass function for the lenses involved in 
these plots include main-sequence stars, brown dwarfs, but also white dwarfs,
neutron stars and black holes, which may have large masses without violating 
the lens flux limit constraint. For the MS and BD stars, we adopt the 
following slopes of the present-day mass function $dN/d\ln(M)$: $-0.3$ between 
0.03 and 0.7 $M_\odot$, $-1.0$ between 0.7 and 1.0 $M_\odot$, and $-4.0$ above. 
For the remnants, we adopt gaussian distributions, whose mean value, standard 
deviation, and fraction of total mass with respect to MS and BD stars below 
$1.0\ M_\odot$ are given in Table~\ref{tab:massdist}.

\begin{deluxetable}{lllr}
\tablecaption{Mass distribution of remnant stars (white dwarfs, neutron stars 
and black holes), with respect to main sequence stars and brown dwarfs below 
$1.0\ M_\odot$\label{tab:massdist}}
\tablewidth{0pt}
\tablehead{Remnant & $<M>$ & $\sigma$ & ratio}
\startdata
WD & 0.6  & 0.07 & 22/69 \\
NS & 1.35 & 0.04 & 6/69 \\
BH & 5.0  & 1.0  & 3/69
\enddata
\end{deluxetable}

For details about the choice of these numbers, please refer to \citet{gould00a}.

\begin{figure}[th]
\epsscale{1.0}
\plottwo{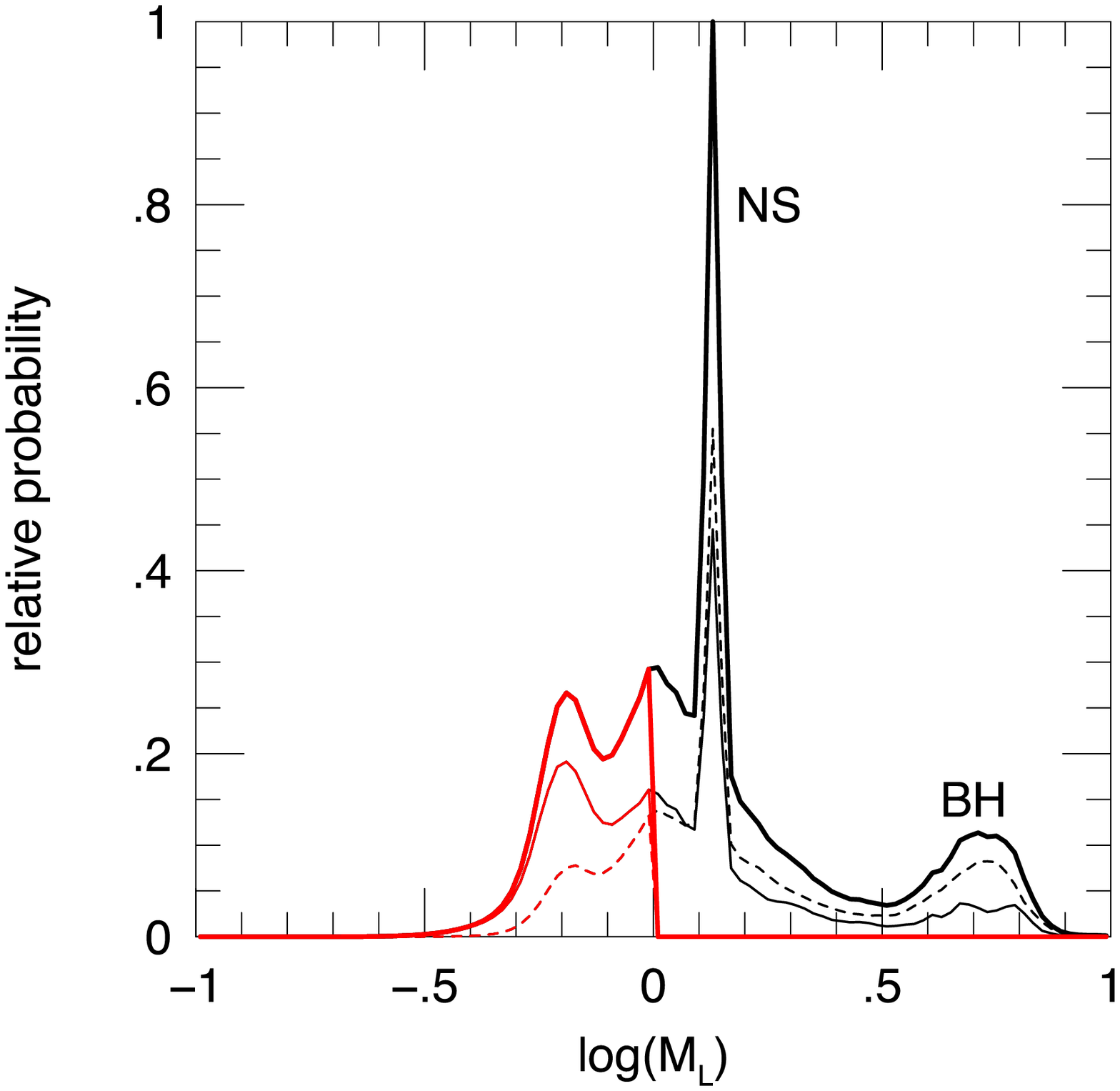}{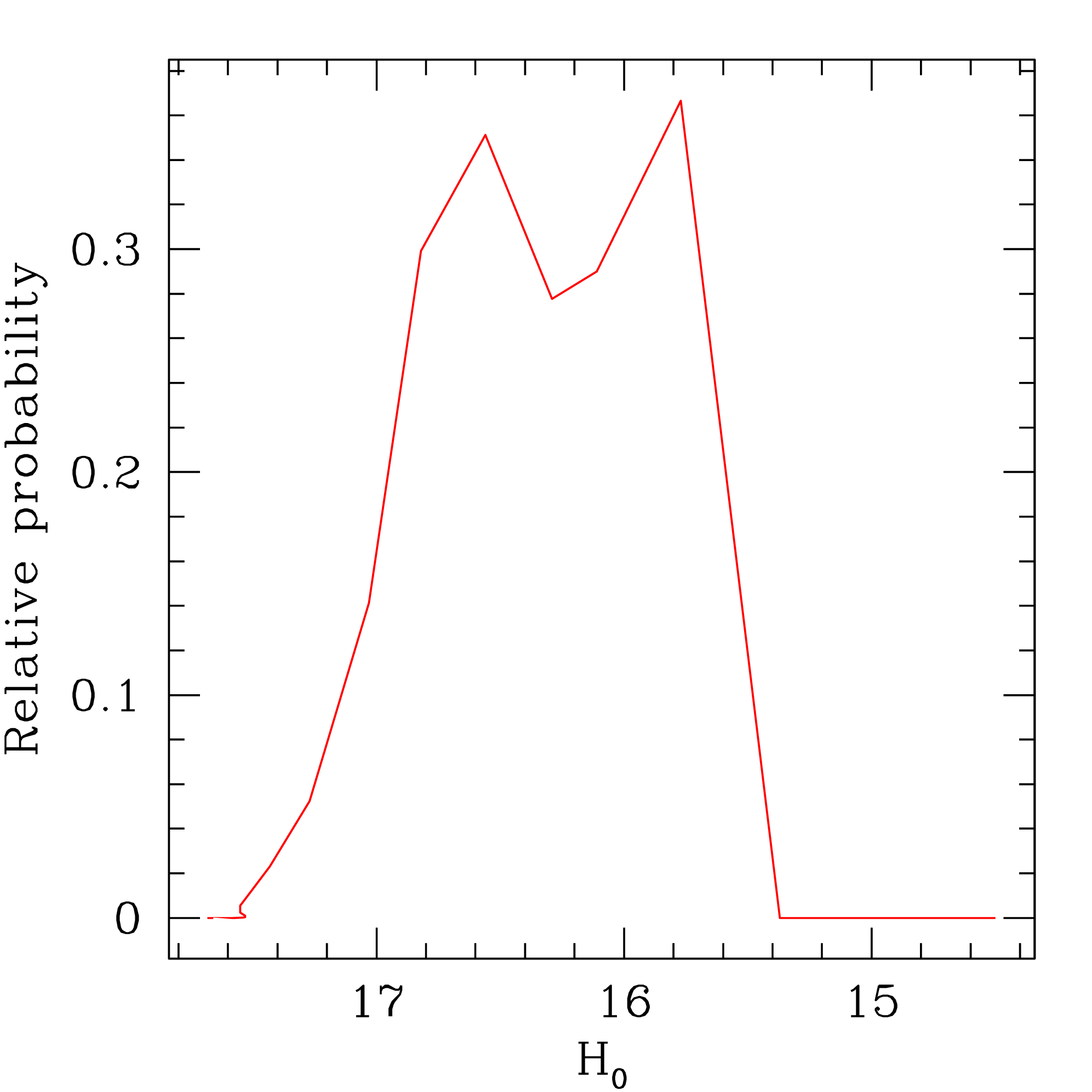}
\caption{\label{fig:1Dplots1}
Distribution of lens mass and lens flux in $H$-band (from left to right) 
computed from the post-bayesian analysis. In the left diagram, the black curves 
show the full mass function, while the red curves show the mass function 
truncated at $1\ M_\odot$. The solid curves represent the $u_0>0$ solution, 
while the dashed curves correspond to the $u_0<0$ one. The bold solid curves 
are the sum of the two other curves. In the right diagram, the curve 
corresponds to the red bold solid curve of the mass distribution.}
\end{figure}

\begin{figure}[th]
\epsscale{1.0}
\plottwo{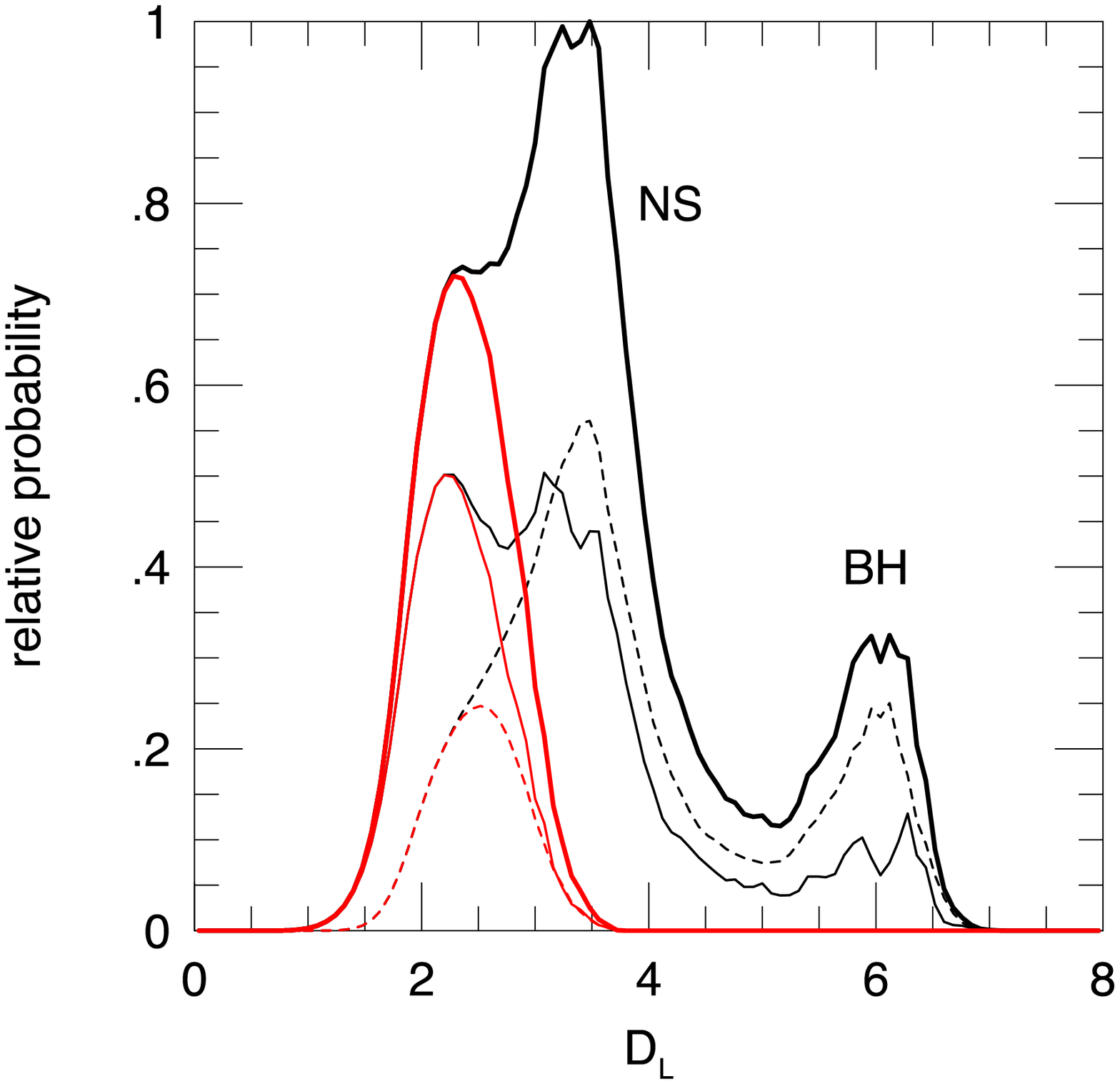}{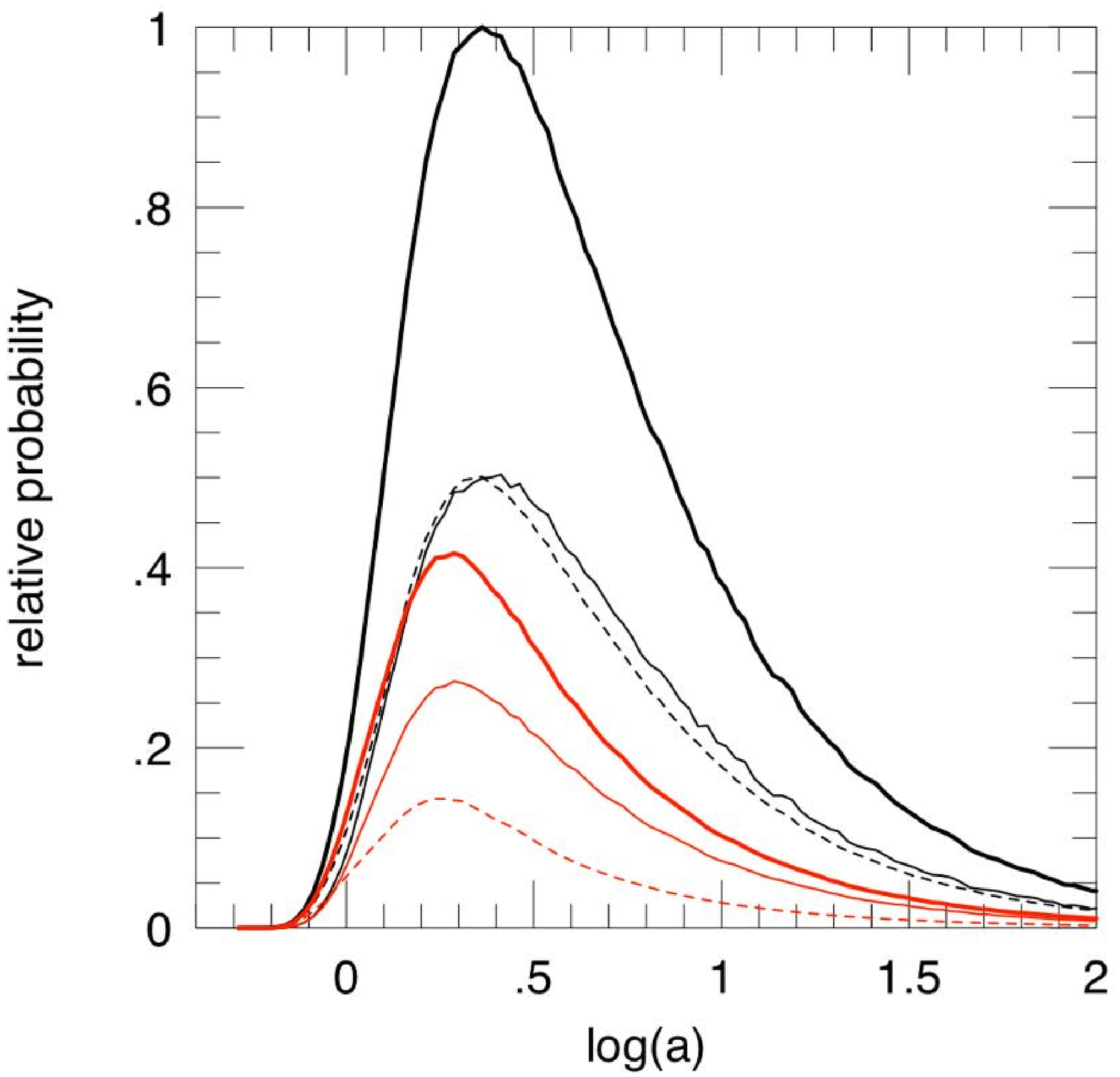}
\caption{\label{fig:1Dplots2}
From left to right: distribution of lens distance (in kpc) and semi-major 
axis of the planet orbit (in AU), computed from the post-bayesian analysis. 
For details, see the description of Figure~\ref{fig:1Dplots1}.}
\end{figure}

In each diagram (see Figure~\ref{fig:1Dplots1} for mass and flux, and 
Figure~\ref{fig:1Dplots2} for distance and semi-major axis), the black curves 
show the full mass function, while the red curves show the mass function 
truncated at $1\ M_\odot$. For MS stars, this limit is imposed by the lens flux 
constraint, and will be refined once we obtain the adaptive optics photometry 
of the individual stars in the field. WD at this mass are extremely rare; 
Jovian planets around pulsars (NS) have not been found, despite very extensive 
searches; and super-Jupiter planets orbiting BH are a priori unlikely. 

Let us first consider the lens mass distribution: the no-flux-limit (black) 
curve shows a huge spike at expected NS position and a smaller bump 
corresponding to BH. Note that for these bumps, the $u_0<0$ solution dominates, 
despite its $\chi^2$ handicap. This is because the Galactic model very strongly 
favors distant lenses, primarily because of the volume factor, and this 
overwhelms the modest preference of the light curve for nearby lenses. 
Because $\theta_{\rm E}$ is roughly fixed, these distant lenses are massive. 
This preference is much stronger in the $u_0<0$ solution, which can be seen in 
its rapid rise beginning at $\log M=-0.3$. Note that the WD peak (at 
$\log M=-0.22$) is clearly visible, especially in the $u_0>0$ solution.

The lens distance distribution basically looks at this same situation from the 
standpoint of distance. The new notable feature is that both MS and NS peaks 
are in the disk, while the BH bump is in the Bulge. And the semi-major axis 
distribution peaks at about 2-3 AU.

>From these diagrams, we can estimate a most probable value of lens mass, 
distance and semi-major axis, and an asymmetric standard deviation read at 50\% 
of the distribution corresponding to the red bold curves. We get a star and 
planet mass of $M_*=0.67^{+0.33}_{-0.13}\ M_\odot$ and 
$m_p=1.5^{+0.8}_{-0.3}\ M_{\rm JUP}$, respectively, at a distance of 
$D=2.3\pm0.6$~kpc, and with a semi-major axis of $a=2^{+3}_{-1}$~AU.

As a final note, it could be said that more complex models are worth exploring:
the geometry of the caustic crossing, where the source passes close to the 
three-cusps tail of the caustic, is extremely sensitive to a third body (second
planet or binary companion to the lens star). A similar geometry where two 
planets were detected is described in \citet{gaudi08,bennett-ogle109}. These 
models could be investigated in a forthcoming paper, once we get the lens flux 
measurement from adaptive optics.

\acknowledgments 
We acknowledge the following sources of support: Creative Research Initiative 
Program (2009-0081561) of National Research Foundation of Korea (CH); grants
JSPS20340052 and JSPS22403003 for MOA; Czech Science Foundation grant 
GACR P209/10/1318; the French Polar Institute (IPEV) and the Italian Antarctic 
Programme (PNRA) for the logistics and data transmission at Concordia. OGLE 
project  has received funding from the European Research Council under the 
European Community's Seventh Framework Programme (FP7/2007-2013) / ERC grant 
agreement no. 246678. Part of the computer work was performed using HPC 
resources from CALMIP (Grant 2011-P1131).

E.~Bachelet gratefully acknowledges the Chungbuk National University for a 
one-month stay where most of this work was prepared. B.S.~Gaudi and 
A.~Gould acknowledge support from NSF AST-1103471. B.S.~Gaudi, A.~Gould, and 
R.W.~Pogge acknowledge support from NASA grant NNG04GL51G. Work by J.C.~Yee is 
supported by the National Science Foundation Graduate Research Fellowship under 
Grant No.~2009068160. T.C.~Hinse acknowledges support from the KRCF Young 
Scientist Research Fellowship Program in South Korea. The PLANET collaboration 
acknowledges the financial support of ANR HOLMES and PNPS grants. ASTEP was 
financed through the help of ANR, IPEV, CNRS, Observatoire de la C\^ote d'Azur. 

\appendix

\section{Observations at Dome~C}\label{app:astep}
ASTEP~400 is a 40cm Newtonian telescope installed at the Concordia base, 
located on the Dome C plateau in Antarctica \citep{daban11}. Although the aim 
of the project concerns transiting planets, the ability to observe 
near-continuously during the antarctic winter and the excellent weather on site
\citep{crouzet10} imply that the telescope can usefully complement 
microlensing observations from other sites, even though the declination of the 
fields and their crowding make the analysis difficult.  

The observations with ASTEP 400 started on August 12, 2010, 15:27 UT after the 
first alert was sent by email and phone to Concordia. On August 14, 23:59 UT, 
the observations were stopped because the magnification had become too small 
for useful observations. The weather conditions were excellent. However, the 
seeing conditions were poor (~3-4"), mostly due to the low-declination of the 
field and location of the telescope on the ground. The 229MB of data 
corresponding to 500x500 cropped images were transmitted to Nice through 
satellite connexion around August 17-18 for an in-depth analysis. 

Although the data of this run were not good enough for being used in this study,
this pioneering test will serve for improving the thermics of the acquisition
system and get higher quality future observations of microlensing targets.

\section{Difference Image Analysis using pySIS}\label{app:pysis}
The pySIS3.0 difference image analysis package is fully described in 
\citet{albrow09}. It is based on the original ISIS package 
\citep{alard98,alard00} but the kernel used to transform the reference image to
the current image before subtraction is no longer analytic, but numerical. This
allows dealing with images whose PSF cannot be assimilated to a sum of gaussian
profiles. The numerical kernel has been introduced in image subtraction by 
\citet{bramich08}. One of the regular problems encountered in difference image 
analysis is the choice of the best possible reference image. For astrometry, 
the best seeing image is generally a good choice, if the sky background is not 
too high. For photometry, it is our experience that stacking good images 
improves the result, but only if these images have been acquired during a short
time slot, to avoid light variations of the target and slow variations due for 
instance to small changes in the flat-field. In order to choose good reference 
images, we use a suite of Astromatic software \citep{bertin11}, namely 
SExtractor and PSFex. SExtractor builds catalogues of sources from all images 
of a given telescope, with their characteristics, and PSFex derives a model of 
the PSF of these images, from which we extract a few numbers to estimate the 
image quality (seeing, ellipticity, number of stars). This allows an almost 
automatic selection of the templates, with a final verification by eye to check
the selected images.

Once the images have been subtracted using these templates, the photometry of 
the target is done by iteratively centering the PSF on the light maximum close 
to the center of the image. This method enables the photometering of faint stars
in the glare of nearby much brighter stars.

\section{Noise Model and Error Rescaling}\label{app:rescaling}

The standard procedure of rescaling error bars so that the $\chi^2/{\rm dof}$ of
each telescope data set is of order unity is acceptable if the resulting error 
bars roughly correspond to the dispersion of the data points at a given time 
for this data set. There is therefore an interplay between finding the correct 
model and rescaling the error bars, because a too large rescaling factor reduces
the constraint from a given data set and allows the model to shift from the 
correct one. Our procedure has been to use rescaling factors which look 
plausible given the telescope size and site quality, trying to get a 
$\chi^2/{\rm dof}$ of order unity only if the dispersion in successive data 
points is well reproduced by this rescaling scheme. This generally involves two
parameters to modify the original photometric error bars $e_{\rm ori}$. One is a 
minimal error $e_{\rm min}$ to reproduce the dispersion of very bright (or in 
this case, highly magnified) sources due to fundamental limitations of the 
photometry, such as flat fielding errors. The other is the classical 
multiplicative rescaling factor $f$. The adopted formula is:
\begin{equation}
e = f \sqrt{e_{\rm ori}^2 + e_{\rm min}^2}
\end{equation}

\begin{figure}[th]
\epsscale{1.0}
\plottwo{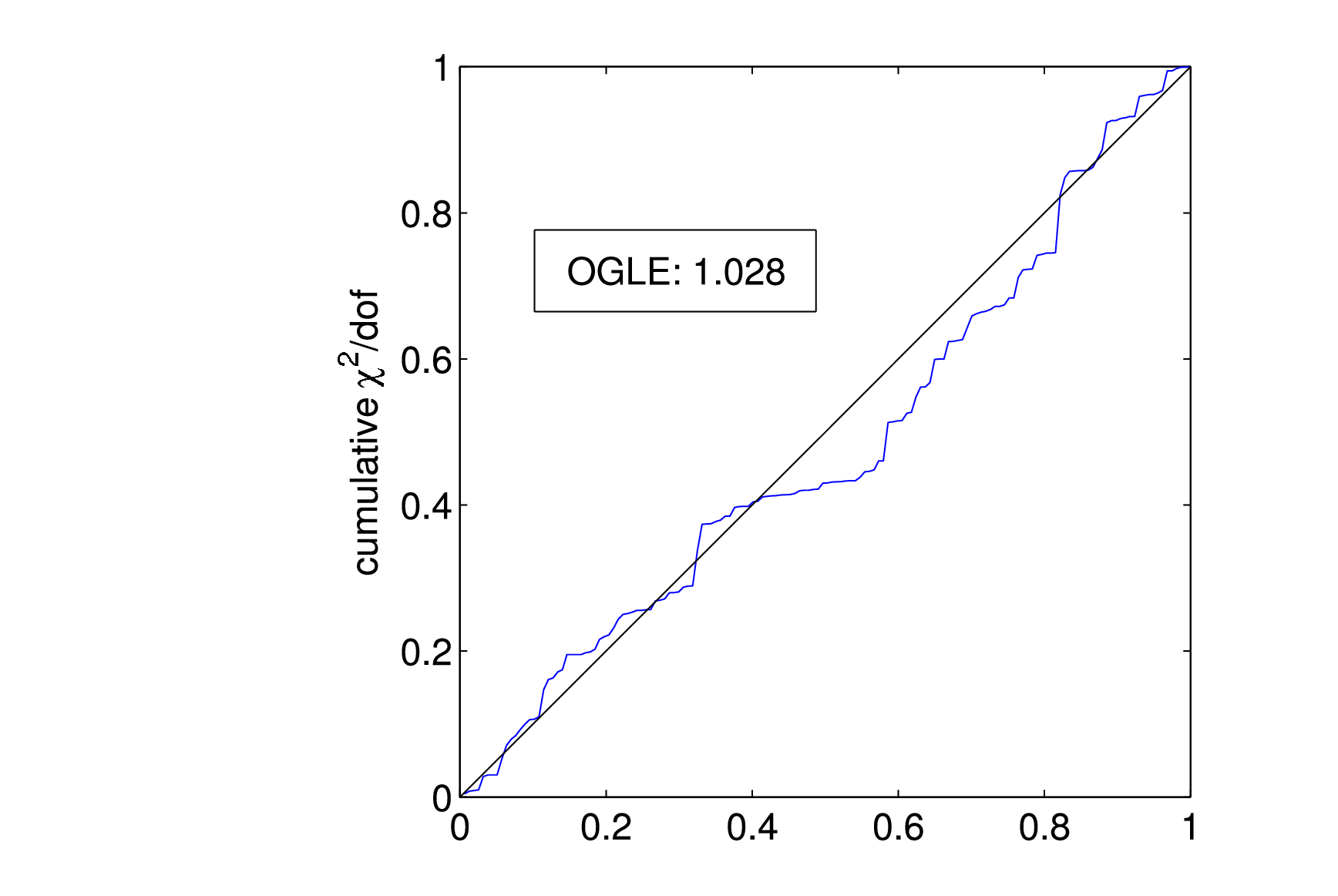}{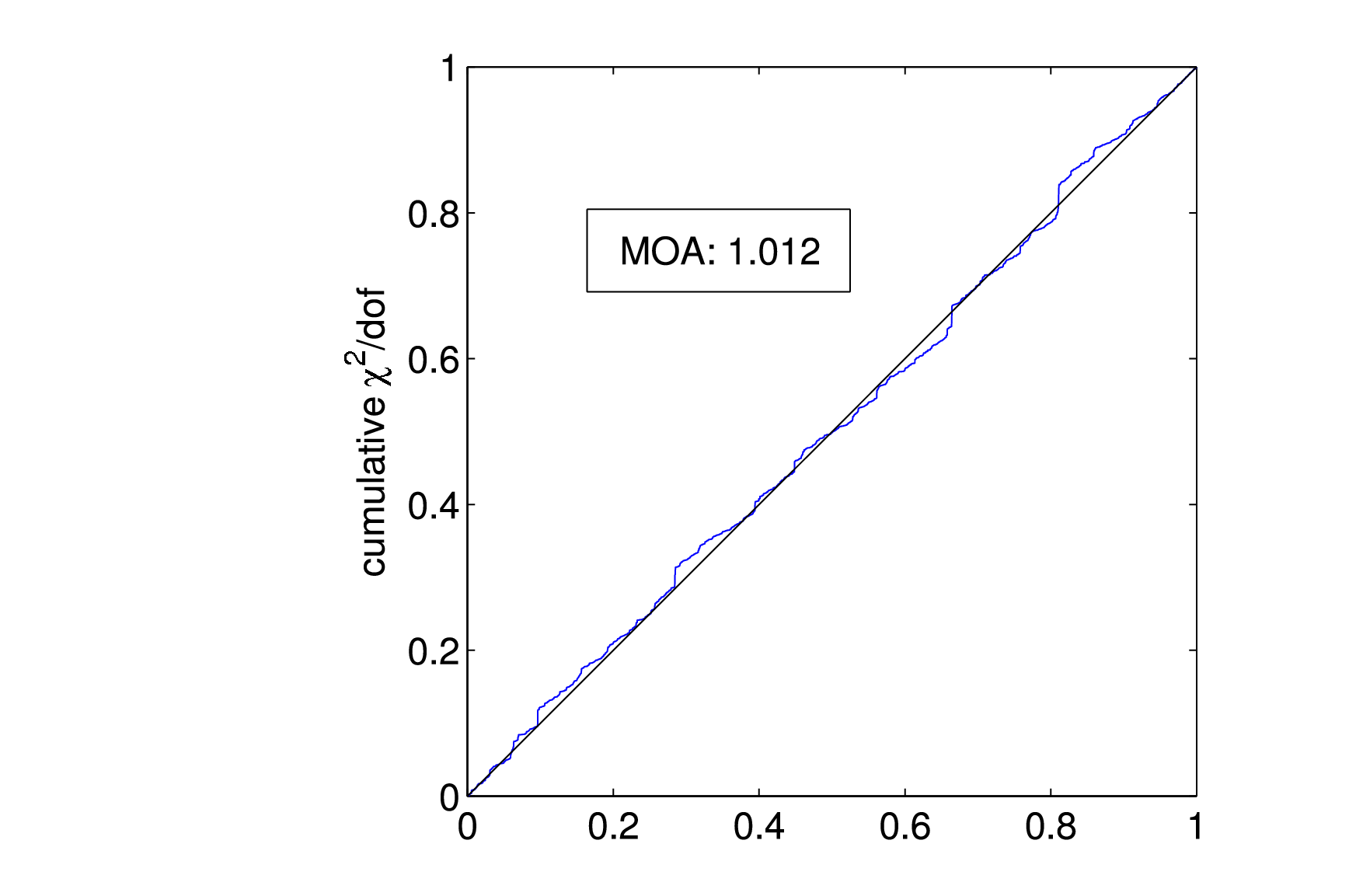}
\caption{\label{fig:cumul}
Cumulative $\chi^2$/dof distribution for rescaled error bars in the case of
OGLE photometry (left) and MOA photometry (right).}
\end{figure}

The balance between both rescaling factors is given by comparing the cumulative
distribution of $\chi^2/{\rm dof}$, ordered from magnification given by the 
model, to a standard cumulative distribution for gaussian errors. An example 
for OGLE and MOA distributions is given in Figure~\ref{fig:cumul}. 
Table~\ref{tab:two} gives the adopted values of both parameters for each 
telescope. It must also be noted that for some amateur telescopes, data were 
binned before this rescaling process. Finally, after the initial modeling was 
conducted, it was realized that three data sets could not accommodate the 
condition $\chi^2/{\rm dof} \sim 1$ (Auckland and VLO) or a positive source 
flux (Perth). They were therefore removed and the final models are based on 16
data sets.


\section{Detailed Treatment of Limb-darkening Corrections}\label{app:lldc}
It is generally difficult to treat the limb-darkening effect accurately, because
it varies from one telescope data set to the other, primarily due to the 
different photometric bands involved. Second-order effects include 
hardware-specific variations of the spectral response (filters, CCDs), and for 
very broad bands, the atmospheric and interstellar extinctions. We adopt the 
linear limb-darkening approximation using the formalism given by 
Equation~\ref{eqn:eq4}, introduced by \citet{albrow99} and \citet{an02}. This 
is a common formalism in the microlensing community, but the more widely used 
formalism is based on the following equation
\begin{equation} \label{eqn:eqd1}
I_{\lambda}(\cos{\phi}) = I_{\lambda}(1) \times \left[ 1- u_{\lambda} (1-\cos{\phi}) \right]
\end{equation}
where $I_{\lambda}(1)$ is the specific intensity at disk center and $u_{\lambda}$ 
is the linear limb-darkening coefficient. We compute the values of $u_{\lambda}$ 
from a stellar atmosphere model from the Kurucz ATLAS9 grid \citep{kurucz93} 
using the method described by \citet{heyrovsky07}. As we converted the 
$u_{\lambda}$ values to our LDC parameter $\Gamma_{\lambda}$, we give the 
$u_{\lambda}$ values in Table~\ref{tab:two}, and recall the conversion relation
\begin{equation}
\Gamma_{\lambda} = \frac{2 \, u_{\lambda}}{3-u_{\lambda}}
\end{equation}

\begin{deluxetable}{lccccrllc}
\tablecaption{Data description for each telescope\label{tab:two}}
\tablewidth{0pt}
\tablehead{Observatory & Band & $u_{\lambda}$ (uncor.) & $u_{\lambda}$ (ext. cor.) & Binning & N & $f$ & $e_{\rm min}$ & Photometry}
\startdata
OGLE     & $I$   & 0.4399 & 0.4363 &   &  157 & 2.4  & 0.010 & OGLE DIA \\
CTIO     & $I$   & 0.4395 & 0.4375 &   &  133 & 1.5  & 0.010 & DoPhot \\
CTIO     & $V$   & 0.5931 & 0.5908 &   &   21 & 2.0  & 0.015 & DoPhot \\
Auckland\tablenotemark{a} & W12   & 0.5016 & 0.4843 & Y &    5 & 1    & 0.003 & DoPhot \\
FCO      & unfiltered  & 0.5550 & 0.5150 & Y &    4 & 1.7  & 0.003 & DoPhot \\
FTS      & $i'$  & 0.4547 & 0.4524 &   &   68 & 3.0  & 0.002 & pySIS \\
Kumeu    & W12   & 0.5016 & 0.4852 & Y &   17 & 3.5  & 0.003 & DoPhot \\
Perth\tablenotemark{a} & $I$   & 0.4325 & 0.4297 &   &   26 & 3.0  & 0.003 & pySIS \\
FTN      & $i'$  & 0.4547 & 0.4524 &   &   72 & 3.8  & 0.006 & DanDIA \\
Possum   & W12   & 0.5204 & 0.5031 &   &   77 & 5.4  & 0.003 & pySIS \\
SAAO     & $I$   & 0.4261 & 0.4217 &   & 2364 & 2.2  & 0.003 & pySIS \\
SAAO     & $V$   & 0.6114 & 0.6086 &   &    3 & 1.2  & 0.003 & pySIS \\
VLO\tablenotemark{a} & unfiltered  & 0.5442 & 0.5022 & Y &    3 & 1    & 0.003 & DoPhot \\
Wise     & unfiltered  & 0.5522 & 0.5131 & Y &    9 & 0.8  & 0.003 & DoPhot \\
Canopus  & $I$   & 0.4355 & 0.4335 &   &   50 & 1.5  & 0.010 & pySIS \\
Danish   & $I$   & 0.4394 & 0.4370 &   &  167 & 2.2  & 0.003 & pySIS \\
MOA      & red   & 0.4754 & 0.4694 &   & 3985 & 1.05 & 0     & MOA DIA \\
LT       & $i'$  & 0.4543 & 0.4513 &   &   41 & 4.7  & 0     & DanDIA \\
Monet N  & $I$   & 0.4414 & 0.4392 &   &  130 & 4.3  & 0.004 & pySIS
\enddata
\tablecomments{a: This data set was not used in the final models}
\end{deluxetable}

For the source characteristics, we adopt the spectroscopic result 
($T_{\rm eff}=6000$~K, $\log g=4.0$, [Fe/H]=0.0). We thus get an estimate of the 
limb-darkening effect for each data set. Table~\ref{tab:two} shows the adopted 
values of $u_{\lambda}$ for each individual telescope and band combination, 
both with and without interstellar extinction correction, together with the 
final number of data points after rejection of possible outliers, and details 
about the image reduction process. As can be seen, these coefficients vary even 
if the photometric band is supposedly the same. See for instance the Cousins 
$I$-band and SDSS $i'$-band, or the Wratten \#12 filter (W12) used at some 
amateur telescopes to mimic an $R$ band. More details about the derivation of 
these values for each telescope can be found in 
\citet{heyrovsky07,fouque10,muraki11}.

We have also computed the effect of interstellar extinction on the LDCs, as it
is not a priori negligible: we typically find differences of a few $10^{-3}$ 
for a data set with a filter and a few $10^{-2}$ for unfiltered ones. The 
magnitude of the effect could then be neglected for filtered data sets, but not
for unfiltered ones or very broad band filters. To give an idea, we compare its 
effect to the uncertainty of our spectroscopic determination of the effective 
temperature of the source, about 150~K. For the OGLE $I$-band, the difference 
between LDCs with and without interstellar extinction correction for an 
adopted extinction of 1.2 mag corresponds to a shift in effective temperature 
of 40~K, while for the Possum broad W12 filter, it corresponds to about 200~K, 
and for an unfiltered data set, it gives a shift of about 500~K.

\section{Detailed analysis of the Kepler constraint}\label{app:jacobi}

\begin{figure}[th]
\epsscale{0.8}
\plotone{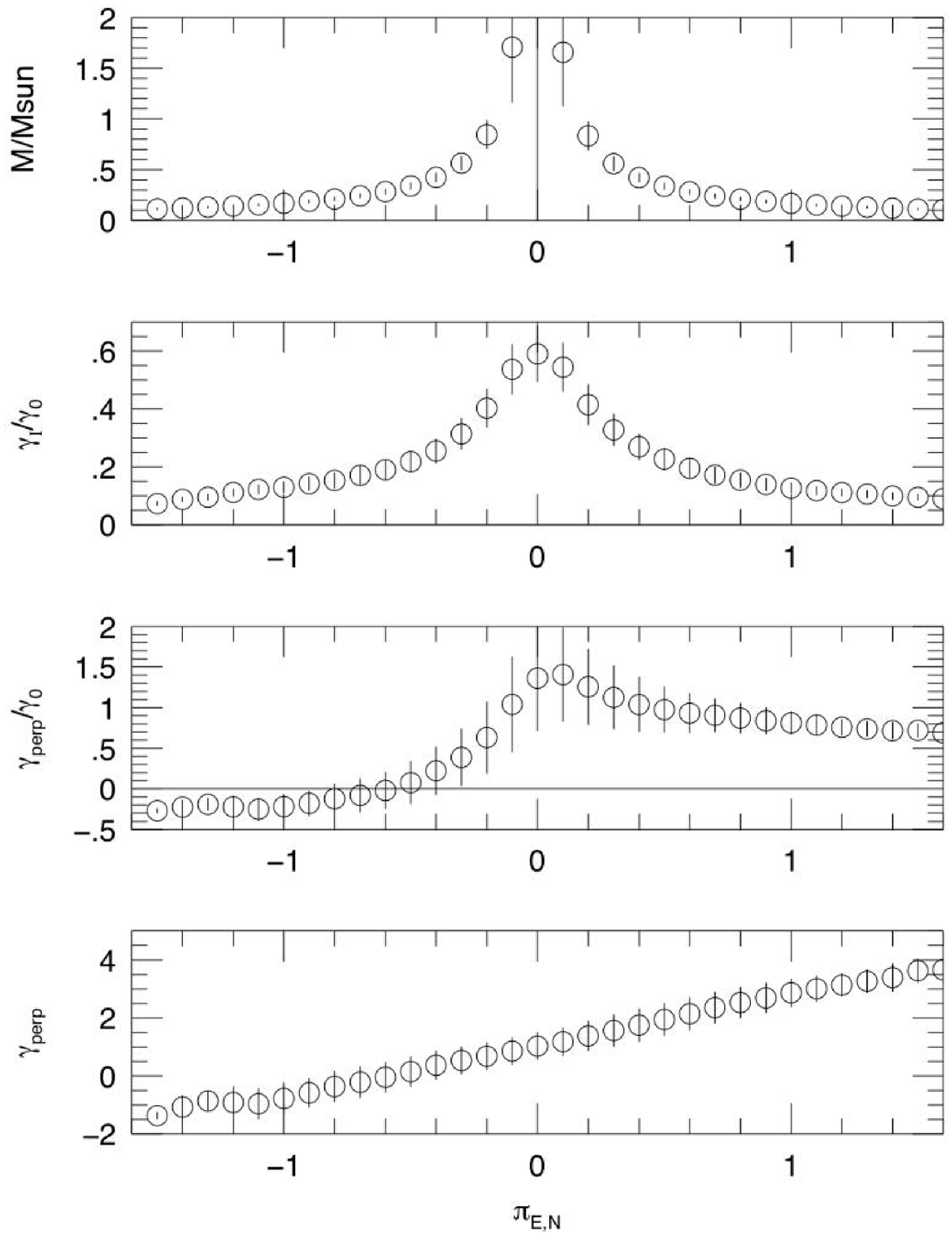}
\caption{\label{fig:anal}
Binned lens characteristics as a function of MCMC variable $\pi_{\rm E,N}$.
Bottom panel shows that $\pi_{\rm E,N}$ is highly correlated with $\gamma_\perp$, 
the component of orbital motion perpendicular to the projected planet-star 
axis, as found earlier by \citet{batista11}. This is responsible for the poor 
determination of $\pi_{\rm E,N}$ in Figure~\ref{fig:bayes1}(a). Top panel: 
highest masses are associated with low $\pi_{\rm E,N}$. Third panel: these are 
disfavored because $\gamma_\perp/\gamma_0$ tends to saturate its physical limit 
(unity). Second panel: relatively high masses (with $\pi_{\rm E,N}<0$) are
favored because $\gamma_\parallel/\gamma_0$ (as well as $\gamma_\perp/\gamma_0$) 
hover near 0.5, a value favored by the virial theorem.}
\end{figure}

Figure~\ref{fig:anal} shows several quantities plotted against
$\pi_{\rm E,N}$.  In each case the mean and standard deviation of all
chain links within a given $\pi_{\rm E,N}$ bin are calculated. The
bottom panel shows the behavior of $\omega$ ($=\gamma_\perp$).  Since
$\pi_{\rm E,N}$ is very similar to $\pi_{\rm E,\perp}$, this reflects
the degeneracy between the perpendicular components of Earth orbital
motion (parallax) and lens orbital motion, which is analyzed in some
detail by \citet{batista11} and \citet{skowron11}. In the present
case, the correlation is quite tight. The top panel shows the lens
mass $M=\theta_{\rm E}/\kappa\pi_{\rm E}$. Since $\theta_{\rm E}$ is
nearly constant for different links in the chain, the mass scales
$M\sim\pi_{\rm E}^{-1}$ and so is peaked near $\pi_{\rm E,N}=0$ where
$\pi_{\rm E}$ is near its minimum (see Figure~\ref{fig:bayes1}).

It is the two middle panels that enable one to understand why the orbital 
Jacobian favors relatively high masses. These show, respectively 
$\gamma_\parallel/\gamma_0$ and $\gamma_\perp/\gamma_0$, where
\begin{eqnarray}
\gamma_{\parallel} & = & s^{-1} \ \dot{s}, \\
\gamma_{\perp} & = & \omega, \\
\gamma_0^2 & = & \frac{8\pi^2}{\kappa\,s^3\pi_{\rm E}}\,
\theta_{\rm E}\,\biggl(\pi_{\rm E} + \frac{\pi_{\rm s}}{\theta_{\rm E}}\biggr)^3
\label{eqn:gamma0}
\end{eqnarray}

Note that $(\gamma_\perp^2+\gamma_\parallel^2)/\gamma_0^2$ is the ratio of the 
so-called projected kinetic to projected potential energy, which has a strict 
upper limit of unity for bound orbits. We do enforce this limit, but the main 
impact of the orbital motion is more subtle. First look at 
$\gamma_\perp/\gamma_0$. In the region away from $\pi_{\rm E,N}\sim 0$, we have 
both $\gamma_0\propto \pi_{\rm E}$ (Equation~\ref{eqn:gamma0}) and (very 
roughly) $\gamma_\perp \propto \pi_{\rm E}$ (bottom panel). Hence, 
$\gamma_\perp/\gamma_0$ is approximately constant in these two regimes. Much of 
the $\pi_{\rm E,N}>0$ region is at or above the physical limit 
$\gamma_\perp = \gamma_0$, which is why these values are disfavored in 
Figure~\ref{fig:bayes1}(c).  

Now examine $\gamma_\parallel/\gamma_0$. By itself, $\gamma_\parallel$ does not 
vary much with $\pi_{\rm E,N}$, so the form of the structure is basically just 
$\gamma_0^{-1}$, which scales $\propto \pi_{\rm E,N}^{-1}$ away from zero. The 
point is, however, that the overall scale (which is set by the measurement of
$\gamma_\parallel$) is small, so that except near $\pi_{\rm E,N}\sim 0$, 
$\gamma_\parallel/\gamma_0$ is extremely close to zero. Naively, this would seem 
to be disfavored by the virial theorem, but how does the Jacobian ``know'' 
about this? For simplicity of exposition let us consider circular orbits. For 
these, $\gamma_\parallel/\gamma_0=0$ implies that the planet is exactly in the 
plane of the sky: either the orbit is exactly face-on (so this is always true), 
or the orbit just happens to be passing through the plane of the sky at the 
time of the event peak. The Jacobian is ``unhappy'' about either alternative
because there is very little Kepler-parameter space relative to chain-variable 
space at such orbital configurations.

Finally, note that in the immediate neighborhood of $\pi_{\rm E,N}\sim 0$ (where 
$M$ reaches its highest values), $\gamma_\perp$ is frequently at or above its 
physical limit ($\gamma_0$), which is exacerbated by the relatively high 
values of $\gamma_\parallel$. This is reponsible for the modest suppression of 
extremely low parallaxes (high masses) in Figure~\ref{fig:bayes1}. Thus, both 
Kepler and Galactic+flux priors separately predict a relatively high lens mass, 
near the limit of what is permitted by the flux constraint.

\end{document}